\documentclass[aps,prl,showkeys,groupedaddress,footinbib,longbibliography,reprint,floatfix]{revtex4-2}

\usepackage[dvipsnames]{xcolor}
\usepackage{amsmath}
\usepackage{amssymb}
\usepackage{bm}
\usepackage{dcolumn}
\usepackage{graphicx}
\usepackage{times}
\usepackage{hyperref}
\usepackage[normalem]{ulem}
\usepackage{multirow}
\usepackage{lipsum}

\hypersetup{
colorlinks=true,
linkcolor=RoyalBlue,
citecolor=RoyalBlue,
urlcolor=RoyalBlue,
}

\newcommand{\angstrom}{\textup{\AA}}
\renewcommand{\parallel}{\mathrel{/\mskip-4mu/}}

\begin{document}

\title{Anomalous Scaling Laws of Dispersion Interactions in Anisotropic Nanostructures}

\author{Hui Pan}
\email[]{panhui@nus.edu.sg}
\affiliation{Department of Physics, National University of Singapore, Singapore 117551,
Republic of Singapore
}

\author{Yuhua Ren}
\affiliation{Department of Physics, National University of Singapore, Singapore 117551,
Republic of Singapore
}

\author{Jian-Sheng Wang}
\email[]{phywjs@nus.edu.sg}
\affiliation{Department of Physics, National University of Singapore, Singapore 117551, Republic of Singapore
}

\date{\today}

\begin{abstract}
The van der Waals (vdW) dispersion interaction between two finite neutral objects typically follows the standard nonretarded $d^{-6}$ law. Here, we reveal an anomalous $d^{-10}$ scaling law between nanostructures with strong geometric or electric anisotropy, driven intrinsically by symmetry-restricted plasmon interactions. At finite anisotropy ratios, a scaling crossover from $d^{-10}$ to $d^{-6}$ occurs due to plasmon mode competition, marked by a finite critical separation. Furthermore, we demonstrate tunability of interlayer vdW forces in two-dimensional materials with strong in-plane electronic anisotropy. By pushing the conventional lower bound of vdW scaling laws, these findings open new opportunities for tailoring nanoscale forces, with potential applications in low-stiction nanomechanical devices, vdW superstructure assembly, metamaterials, and molecular simulations.
\end{abstract}

\maketitle

Dispersion interactions, known as van der Waals (vdW) attractions between neutral atoms, molecules, or macroscopic objects, arise from long-range correlations between fluctuating charges and are ubiquitous in condensed matter systems~\cite{parsegian2005van,French2010,Stone2013,Woods2016}.
Understanding and controlling these interactions at the nanoscale not only deepens insight into fluctuation-induced phenomena in quantum materials~\cite{Buhmann2012-2,Komnik1998,Tse2007,Volokitin2011,Basov2016,Venkataram2020,Hauseux2022}, but also underpins key applications in nanomechanical systems~\cite{Chan2001,DelRio2005}, self-assembly~\cite{Bishop2009}, metamaterials~\cite{Dorrell2020}, molecular simulations~\cite{Mazzola2017}, and functional nanodevices~\cite{Nerngchamnong2013}.
While the commonly used pairwise summation, based on the interatomic $d^{-6}$ vdW law, captures bulk behavior, it breaks down in low-dimensional, especially metallic, systems due to many-body correlations and wavelike plasmon excitations~\cite{Dobson2001,Dobson2006,Tkatchenko2009,Cole2009,Lebegue2010,Sarabadani2011,Tkatchenko2012,Noruzifar2012,Dobson2014How,Dobson2014,Wagner2014,Reilly2015,Ambrosetti2016,Hermann2020}.
These nonadditive effects typically lead to longer-range interactions with unconventional power laws, fueling sustained interest across diverse geometries and materials~\cite{Gould2009,Gobre2013,Stedman2014,Makhnovets2017,Yang2019,Kou2023,Le2024}.
Yet enhancement is not always beneficial; suppressing stiction between nanostructures is crucial for nanomechanical stability~\cite{DelRio2005}, low-friction interfaces~\cite{Filleter2009}, and reactivity control~\cite{Lee2015}.
Tailoring dispersion interactions across a broad range of power laws remains an open challenge~\cite{Kleshchonok2018}, as the scaling almost inevitably reverts to the $d^{-6}$ law at large separations~\cite{Calbi2003}, severely limiting tunability.
Various strategies have enabled scaling crossover, such as waveguides~\cite{Cho1996,Shahmoon2013,Shahmoon2014}, optical excitation~\cite{Beguin2013,Ambrosetti2022,Tasci2025}, and electric fields~\cite{Fiscelli2020,Karimpour2022}. However, the achieved scaling exponents rarely fall below $-6$, let alone without external intervention.

Anisotropic nanostructures exhibit highly directional, tunable microscopic responses, enabling orientation-dependent vdW forces and torques~\cite{Saville2006,Emig2009,Zhang2017,Lu2018,Wang2024,Kou2024}.
In contrast to isotropic systems, plasmons (collective electron oscillations) in anisotropic ones are dimensionally constrained, manifesting as Luttinger liquids~\cite{Shi2015,Wang2022}.
Specific relative orientations can even alter the asymptotic scaling between crossed conducting wires~\cite{Dobson2009,Rodriguez-Lopez2012}.
However, prior studies focus primarily on separations much smaller than the object length.
Recent works reveal substantial vdW torques from charge density fluctuations in twisted anisotropic nanosheets~\cite{Wang2024,Kou2024}.
Unlike the well-explored collinear and parallel cases~\cite{White2008,Misquitta2010,Misquitta2014}, systems with large relative angles and separations exhibit scaling behavior that remains largely unexplored.

\begin{figure}
    \centering
    \includegraphics[width=\columnwidth]{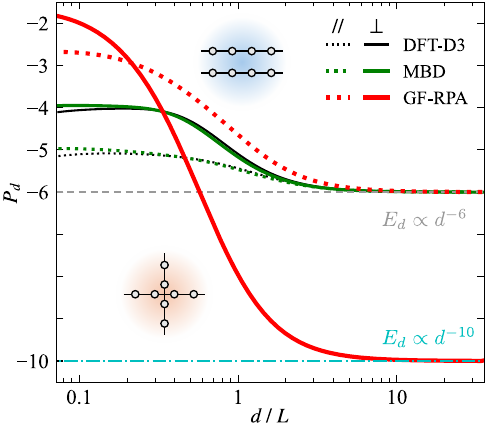}
    \caption{Scaling exponents of dispersion energy between two atomic chains as a function of separation distance $d$ for parallel ($\parallel$, dotted lines) and perpendicular ($\bot$, solid lines) configurations (insets).
    Both chains have length $L = 100 a$ with bond length $a = 1.4~\angstrom$.
    In contrast to the parallel case, the scaling exponent for the perpendicular setup approaches $-10$ rather than $-6$ at large separations, as given by the Green's function method with the random phase approximation (GF-RPA). Results from the many-body (MBD) and density functional (DFT-D3) dispersion methods are also shown.}
    \label{Fig:fig1}
\end{figure}

\begin{figure}
    \centering
    \includegraphics[width=\columnwidth]{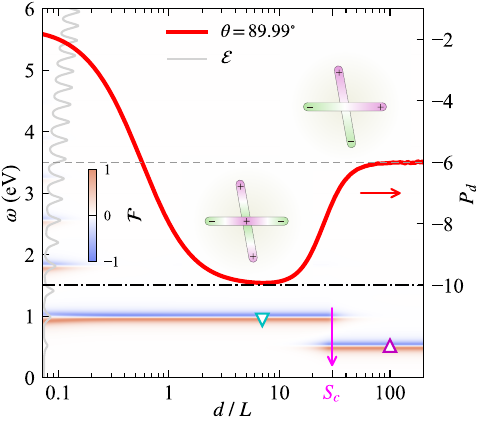}
    \caption{ Crossover of the scaling exponent (red line, right axis) of dispersion energy between two atomic chains of length $L = 100 a$ in a nearly perpendicular configuration with crossing angle $\theta = 89.99^{\circ}$. Resonance peaks in the electron energy loss spectrum ($\mathcal{E}$, gray solid line) correspond to discrete plasmon energies of a single chain. The dominant mode of spectral force ($\mathcal{F}$, background map) transits from even (down triangle) to odd (up triangle) near the critical separation $S_c = 30$. Insets show wavelike charge polarization patterns for the two lowest plasmon modes. }
    \label{Fig:fig2}
\end{figure}

\begin{figure}
    \centering
    \includegraphics[width=\columnwidth]{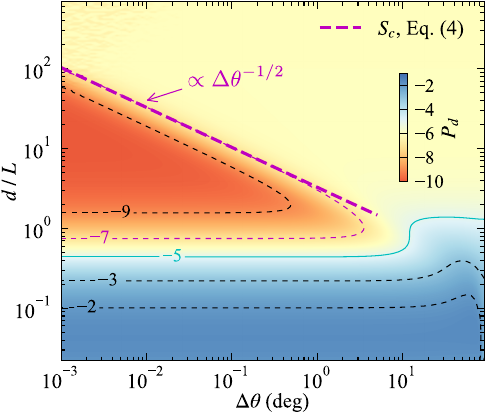}
    \caption{Scaling exponent (contour map) of dispersion energy between two atomic chains crossing at angle $\theta$, plotted as a function of angle deviation $\Delta\theta = \pi/2 - \theta$ and scaled distance $d/L$. The critical separation $S_c$ decays exponentially with $\Delta\theta$ and follows a $\Delta\theta^{-1/2}$ scaling for small $\Delta\theta$, in good agreement with Eq.~\eqref{Eq:critical-separation-llt} (thick dashed line) derived from Luttinger liquid theory. The effective separation region is defined as the distance range where $P_d < -6$ (red area).}
    \label{Fig:fig3}
\end{figure}

\begin{figure}
    \centering
    \includegraphics[width=\columnwidth]{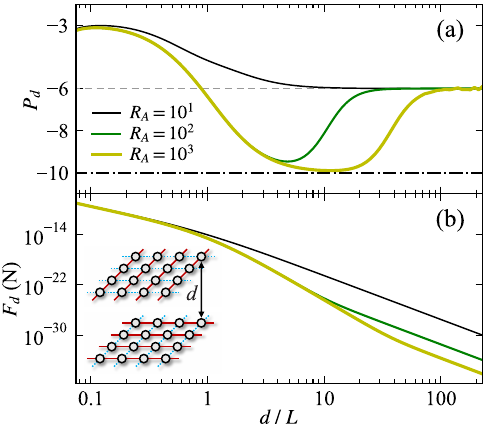}
    \caption{(a) Scaling exponents of dispersion energy and (b) corresponding dispersion forces between two square metal plates (inset) with mutually perpendicular anisotropic axes and side length $L = 30 a$, for various in-plane hopping anisotropy ratios $R_A = t_a / t_b$ of $10$, $100$, and $1000$, where $t_a$ (solid bonding lines) and $t_b$ (dotted bonding lines) are the anisotropic hopping parameters. The unusual $P_d < -6$ scaling and exponent crossover also arise from electronic anisotropy.
    Both the scaling exponent and dispersion force strongly depend on $R_A$ at a given separation distance.}
    \label{Fig:fig4}
\end{figure}

In this Letter, we reveal how the textbook $d^{-6}$ vdW limit can be intrinsically overcome by collective excitations in anisotropic structures without external actions. We consider two crossed carbon chains, as a minimal instance of systems with strong geometric anisotropy such as nanotubes, where plasmon modes exhibit wavelike charge oscillations~\cite{Shi2015,DeVega2016}.
We find that the asymptotic scaling law switches from the standard $d^{-6}$ to an anomalous $d^{-10}$ when the relative orientation changes from parallel to perpendicular  (Fig.~\ref{Fig:fig1}), challenging the conventional knowledge about vdW law. Spectral analyses show that this arises from symmetry-forbidden dipolar plasmon coupling (Fig.~\ref{Fig:fig2}), leaving high-order modes dominant, further confirmed by Luttinger liquid theory. For arbitrary angles, we derive an analytical criterion of the scaling crossover due to mode competition, in agreement with numerical results (Fig.~\ref{Fig:fig3}). More importantly, such anomalous scaling laws also emerge in two-dimensional materials with strong electronic anisotropy, driven by the same mechanism (Fig.~\ref{Fig:fig4}).
These findings open new opportunities for tailoring nanoscale forces, extensible to broader fluctuation physics.

To naturally capture both dispersion interactions and plasmon excitations in metallic systems, we present a microscopic numerical framework based on the nonequilibrium Green's function (NEGF) formalism with the random phase approximation (RPA).
The NEGF has been utilized for exploring thermal radiation and Casimir physics~\cite{Wang2023,Chudnovskiy2023,Pan2024}
, and the RPA has been recognized as the benchmark in evaluating vdW interactions~\cite{Dobson2012}, especially for conductors where plasmon dominates.
In contrast to established methods~\cite{Toulouse2009,Tkatchenko2013,Drosdoff2014,Ramberger2017}, our approach operates in the \emph{real}-frequency domain, enabling spectral analysis of individual plasmon-mode contributions.

The total interaction energy between the transient electric potential $\phi$ and fluctuating charge density $\rho$ reads $E =\operatorname{Tr}\left\langle{\phi\rho}\right\rangle$ , where ``$\operatorname{Tr}$'' denotes spatial integration~\cite{Stone2013}.
Note that $\rho$ and $\phi$ are correlated by the Poisson equation $\rho = v^{-1}\phi$, with $v = 1/4\pi\epsilon_0r$ the bare Coulomb interaction~\cite{jackson1999classical}.
For a system with two objects separated by a distance $d$, the dispersion energy is the separation-dependent part of the total~\cite{Dobson2006}, $E_d = E(d)-E(d\to\infty)$.
The dispersion force on object $\alpha$ is given by the gradient as $\bm{F}_{\alpha} = -\operatorname{Tr}\left\langle{\nabla}\phi\rho\right\rangle_{\alpha}$.
We define the scalar Green's function and self-energy in contour time as $D\left(\tau,\tau'\right) = -i \left\langle \mathcal{T} \phi\left(\tau\right)\phi\left(\tau'\right)\right\rangle/\hbar$ and $\Pi\left(\tau,\tau'\right)=-i\left\langle\mathcal{T}\rho\left(\tau\right)\rho\left(\tau'\right)\right\rangle/\hbar$, respectively, with $\mathcal{T}$ the contour-ordering operator~\cite{Wang2023}.
Accordingly, the dispersion force can be written as
\begin{equation}\label{Eq:disp-force}
\bm{F}_{\alpha} = 2\operatorname{Im}\int_{0}^{+\infty}{\frac{\hbar d\omega}{2\pi}\operatorname{Tr}\left[ {\nabla\!\left(   {D^r\Pi_{\alpha}^r} \right)} \right]\left(2N_B+1\right)},
\end{equation}
where $D^{r}$ gives the dressed Coulomb interaction, satisfying the Dyson equation $D^r = v + v\Pi^r D^r$~\cite{Wang2023}.
Under the RPA, the retarded self-energy yields the bare polarization function $\chi^0=\Pi^r$, given by electron Green's functions~\cite{Toulouse2009}. Here, $N_B\left(\omega,T\right) \!=\! 1/\left[\exp{\left(\hbar\omega/k_B T\right)-1}\right]$ is the Bose-Einstein distribution at frequency $\omega$ and temperature $T$, accounting for thermal fluctuations~\cite{Drosdoff2014}.
Throughout this work, the scaling exponent of dispersion energy between two objects separated by a given distance $d$ is defined as~\cite{Ambrosetti2016}
\begin{equation}
P_{d} \left(d\right) = 1 + \frac{\partial \ln{F_d\left(d\right)}}{\partial \ln{d}},
\end{equation}
where $F_d \equiv \left|\bm{F}_{\alpha}\right|$ denotes the force magnitude on either object, given that $\bm{F}_1 = -\bm{F}_2$.
Additionally, we introduce the electron energy loss spectrum $\mathcal{E}$ to identify discrete plasmon energies~\cite{Settnes2017}, $\mathcal{E}\left(\omega\right) = -\operatorname{Im}\left[\bm{\epsilon}_m^{-1}(\omega)\right]$, where $\bm{\epsilon} = \bm{1} - v \chi_0$ denotes the microscopic dielectric function.
The resonance frequencies corresponding to energy loss peaks indicate plasmon energies, resolvable from the RPA equation~\cite{Li1989,Dobson2006}, $\operatorname{Re}\left[\bm{\epsilon} \left({\omega}\right)\right] = 0$.
Notably, unlike the prior method~\cite{Toulouse2009}, electron corrections are neglected here. Although poor in exact dispersion energy, it successfully captures the power laws at large separations~\cite{Dobson2006,Ambrosetti2016}. More details about obtaining Eq.~\eqref{Eq:disp-force} as well as the formalism can be found in the Supplemental Material~\cite{SM}.

To begin with, we explore the scaling behavior of dispersion interaction between two carbon chains in parallel ({\small $\parallel$}) and perpendicular ($\bot$) scenarios (see insets in Fig.~\ref{Fig:fig1}), as a minimal instance of systems with strong geometric anisotropy, such as nanotubes~\cite{Shi2015,DeVega2016}.
Both chains have length $L = Na$ with $N = 100$ atoms and bond length $a = 1.4$~{\angstrom}.
Figure~\ref{Fig:fig1} shows the scaling exponents $P_d$ as a function of separation distance $d$.
Remarkably, in contrast to the parallel case (red dotted line), the scaling exponent at large separations in the perpendicular setup (red solid line) follows the striking $P_d = -10$ rather than the conventional $P_d = -6$, indicating the breakdown of single dipole approximation even at $d\gg L$~\cite{Calbi2003}.
At small separations, the scaling exponent follows $-1 > P_d > -2$ for perpendicular and $-2 > P_d > -3$ for parallel configurations, consistent with previous analyses~\cite{Dobson2009,Misquitta2010}.
In contrast, neither density functional (DFT-D3, black lines) nor many-body (MBD, green lines) dispersion methods~\cite{Grimme2010,Tkatchenko2012} capture the asymptotic behaviors of both $P_d = -10$ at $d \gg L$ and $-1 > P_d > -2$ at $d \ll L$.
This results from neglecting the wavelike nature of nonlocal plasmon modes in metallic systems~\cite{Nagao2006,Ambrosetti2016,Dobson2023}. Numerical details can be found in the Supplemental Material~\cite{SM}.

In order to uncover the microscopic origin of such unusual $P_d = -10$ asymptotics, we now examine the distance dependence of scaling exponent in the nearly perpendicular scenario by introducing a small angular deviation, e.g., $\Delta\theta = \pi/2 - \theta = 0.01^{\circ}$ (see insets in Fig.~\ref{Fig:fig2}). To identify the contribution from each plasmon mode, we introduce the real-frequency spectral force $\mathcal{F}$, corresponding to the integrand in Eq.~\eqref{Eq:disp-force}, given by
\begin{equation}
\mathcal{F}(\omega) = 2\operatorname{Im}\operatorname{Tr}\left[ \nabla\!\left( D^r \Pi_{\alpha}^r \right)\right]\left(2N_B+1\right) .
\end{equation}
As shown in Fig.~\ref{Fig:fig2}, with increasing separation distance $d$, the dominant plasmon modes of the spectral force (background map) switch from the even (down triangle) to the odd (up triangle), along with the scaling exponent (red line) undergoing a distinct crossover from $-10$ to $-6$, near the critical separation $S_c = d/L=30$. These suggest that the $d^{-6}$ scaling  arises from odd modes, whereas the even modes contributes solely to the $d^{-10}$ decay of dispersion energy.
Allowing for charge polarizations of the two lowest plasmon modes (see insets), in the perpendicular case ($\theta = \pi/2$), the Coulomb interaction between odd modes (right inset, antisymmetric) vanishes exactly due to mirror symmetry, while not for even modes (left inset, symmetric), leading to the asymptotic limit of $P_d = -10$ in Fig.~\ref{Fig:fig1}. However, a small angular deviation $\Delta\theta$ breaks mirror symmetry, resulting in finite contributions from the odd modes and, consequently, the $P_d = -6$ asymptotics at $d \gg L$.
Namely, the scaling crossover along with a finite critical separation results from the competition between even and odd plasmon modes.

We further demonstrate the above landscape analytically by Luttinger liquid theory for one-dimensional plasmons~\cite{Makhnovets2017}.
For two crossed conducting wires of length $L$, the dispersion energy reads $E_d = \hbar\sum\nolimits_{n} {\left(\omega_n^{+} + \omega_n^{-} - 2\omega_n\right)}$, where $\omega_n^{\pm}=\sqrt{\omega_n^2 \pm \mathcal{W}_n}$ is the coupled plasmon frequency due to the interwire Coulomb interaction $\mathcal{W}_n$, while $\omega_n$ is the uncoupled one (i.e., $d\to\infty$) at the $n$-th mode with wave number $q = {\pi n}/ L$.
Since $\mathcal{W}_n \to 0$ when $d \gg L$, the dispersion energy is approximated by the leading nonzero term in $\mathcal{W}_n$, $E_d \approx -\hbar\sum\nolimits_n{{\mathcal{W}_n^2}/4{\omega_n^{3}}}$.
Also, the Coulomb interaction can be expanded in the scaled distance as $\mathcal{W}_n = C_n^{(3)}s^{-3} + C_n^{(5)}s^{-5} + O(s^{-7})$, with $s = \pi d / L$.
As a result, the scaling exponent $P_d = \partial\ln{E_d}/\partial\ln{d}$ at $s \to \infty$ is determined by the leading nonzero term of $\mathcal{W}_n$.
Notably, the $s^{-3}$ term exactly gives the standard $d^{-6}$ vdW law.
After some calculus, the two lowest-order coefficients are found as $C_n^{(3)} = 0$ and $C_n^{(5)} = 3\pi^2(2+\cos{2\theta})/n^2$ for even $n$, and $C_n^{(3)} = 4\cos{\theta}/n^2$, $C_n^{(5)} = 9(8-n^2\pi^2)\cos{\theta}/n^4$ for odd $n$.
Consequently, one can claim two key results: (i) the $s^{-3}$ term is contributed solely by odd modes, and (ii) $C_n^{(3)} = 0$ while $C_n^{(5)} \neq 0$ at $\theta = \pi/2$, consistent with the spectral force analysis in Fig.~\ref{Fig:fig2}. Detailed derivations can be found in the Supplemental Material~\cite{SM}, where a qualitative argument based on the dipole model is provided as well (see Fig.~S2).

It is worth noting that, as the fingerprint of plasmon mode competition, the critical separation $S_c$ strongly depends on the crossing angle $\theta$.
We next explore the dependence of the effective separation region corresponding to $P_d < -6$ on the angular deviation $\Delta\theta = \pi/2 - \theta$.
Figure~\ref{Fig:fig3} shows the scaling exponent as a function of $\Delta\theta$ and the scaled distance $d / L$.
Significantly, the effective separation region (red area) remains observable up to  $\Delta\theta \approx 7^{\circ}$, where the critical separation $S_c$ (upper edge) decays exponentially with $\Delta\theta$, while the lower limit stays around $d / L = 1/2$.
Since the dispersion interaction at small $\Delta\theta$ is dominated by the two lowest modes for $d > L$ (see Fig.~\ref{Fig:fig2}), the exponential dependence of $S_c$ on $\Delta\theta$ can be further identified within Luttinger liquid theory by applying the critical condition $\mathcal{W}_{n=1}^{(3)} = \mathcal{W}_{n=2}^{(5)}$, which simply yields~\cite{SM}
\begin{equation}\label{Eq:critical-separation-llt}
S_c \left({\theta}\right) = \frac{\sqrt{3}}{4} \sqrt{\frac{1}{\cos{\theta}} + 2\cos{\theta}} .
\end{equation}
Here, we have ignored the $\mathcal{W}_{n=1}^{(5)}$ term because of the fact that $\mathcal{W}_{n=1}^{(5)} \ll \mathcal{W}_{n=2}^{(5)}$ for $\theta \to \pi/2$. Note that $\theta \in \left[ 0, \pi/2 \right)$.
As depicted in Fig.~\ref{Fig:fig3}, the $S_c$ from Eq.~\eqref{Eq:critical-separation-llt} (thick dashed line) agrees well with the numerical contour of $P_d = -7$. Since Eq.~\eqref{Eq:critical-separation-llt} is derived from small angular deviations, the disagreement for $\Delta\theta > 1^{\circ}$ results from the finite contributions of high-order modes ($n \geq 3$) to the Coulomb interaction.
In the limit of $\Delta\theta \to 0$, one obtains the scaling relation $S_c \propto (\Delta\theta)^{-1/2}$.
The results here indicate that symmetry-breaking factors, such as twisting, can manipulate intermolecular forces and even power laws in anisotropic structures. Also, Eq.~\eqref{Eq:critical-separation-llt} provides an analytical criterion for experimental realization.

In addition to geometric anisotropy, such anomalous scaling laws and exponent crossover can also emerge in systems with strong electronic or dielectric anisotropy~\cite{Wang2022,Yu2023}.
We demonstrate this by examining the dispersion interaction between two square nanosheets with isotropic geometries but in-plane electron anisotropy, quantized by the hopping ratio $R_A=t_a/t_b$, where the anisotropic axes are mutually perpendicular [see inset in Fig.~\ref{Fig:fig4}(b)].
As expected, Fig.~\ref{Fig:fig4}(a) shows that the dispersion interaction between the sheets without geometric anisotropy also exhibits the unusual $P_d < -6$ scaling behavior, along with an exponent crossover from $P_d = -10$ to $P_d = -6$, driven solely by electronic anisotropy.
Also, the critical separation $S_c$ increases with $R_A$ and tends to diverge as $S_c\to\infty$ for $R_A\to\infty$.
Notably, unlike the mirror symmetry breaking mechanism in anisotropic geometries, it is the finite $R_A$ that gives rise to finite $s^{-3}$ contributions from surface plasmon interactions, thereby resulting in a finite $S_c$.
Furthermore, as shown in Fig.~\ref{Fig:fig4}(b), the dispersion force at specific separations significantly depends on the hopping anisotropy.
For instance, at distance $d = 2 L$, the magnitude of $F_d$ for $R_A = 10$ is about twenty times that for $R_A = 100$.
This dependence enables control of interlayer interactions via in-plane electronic anisotropy,
with the minimal working distance set by the critical separation at the lowest $R_A$, offering prospects for functional molecular devices~\cite{Nerngchamnong2013}.

We conclude by discussing the applicable range, experimental feasibility, and broader relevance of our findings. We remark that the unusual vdW scaling laws of $P_d < -6$ uncovered from the carbon chains applies to quasi-one-dimensional structures, such as nanowires and nanotubes~\cite{Nagao2006,Shi2015}.
Nonetheless, the finite radius results in nonzero $d^{-6}$ contributions from radial polarizations (plasmon modes), making the critical separation dependent on the aspect ratio.
We demonstrate this by examining mutually perpendicular carbon nanotubes (CNTs) (see Fig.~S1 in  the Supplemental Material)~\cite{SM}.
Notably, retardation effects~\cite{Casimir1948} are negligible here, since the unusual power laws emerge in the nonretarded (near-field) regime (refer to Fig.~\ref{Fig:fig2}), i.e., $\lambda_0 = 2 \pi c / \omega_0 > d$, where $\lambda_0$ and $\omega_0$ denote the characteristic optical wavelength and plasmon frequency, respectively.

The experimental realization of our results would be intriguing in itself.
Rydberg~\cite{Beguin2013,Zeybek2023} and ultracold~\cite{Yang2017} atoms provide versatile platforms for exploring vdW interactions and Luttinger liquids.
For material systems, various optomechanical systems have been proposed for detecting piconewton forces~\cite{Intravaia2013}, and atomic force microscopy have been used for layered structures~\cite{Li2019}.
As candidate materials, ultralong gold nanowires and carbon chains ($> 100$ atoms) have been synthesized~\cite{Wang2008,Gao2024}, while homochiral CNTs and ultrathin GaTe with high electric or optical anisotropy ($>10^3$) have been developed~\cite{Wang2019,Zhang2025}.
Additionally, Luttinger-liquid plasmon has been observed in metallic CNTs~\cite{Shi2015} and twisted WTe$_2$~\cite{Wang2022,Yu2023}.
All these advances suggest experimental verification well within reach.
More broadly, the landscape of plasmon–symmetry interplay revealed here transcends the immediate context of vdW forces.
Since retardation effects typically reduce the scaling exponent by one order relative to the non-retarded limit~\cite{Casimir1948,Calbi2003}, extending analysis to Casimir regime would be intriguing.
Beyond metallic systems, the role of microscopic symmetries in dispersion-like interactions between piezoelectric materials~\cite{Le2024Phonon} is also worth exploring.
Likewise, the symmetry-constrained engineering may offer new opportunities for controlling quantum transport~\cite{Komnik1998}, thermal radiation~\cite{Venkataram2020}, and optomechanical responses.

In summary, we have revealed an anomalous $d^{-10}$ power law for dispersion interactions in anisotropic systems, surpassing the textbook $d^{-6}$ vdW law.
Such anomalous scaling originates from symmetry-restricted interactions between wavelike plasmon modes, and exhibits a scaling crossover from $d^{-10}$ to $d^{-6}$ at finite anisotropy ratios due to plasmon mode competition.
In addition to geometric anisotropy, the in-plane electronic anisotropy enables efficient control of interlayer dispersion forces and scaling laws in two-dimensional materials, driven by the same symmetry suppression mechanism.
By extending the accessible power-law range, these findings open new possibilities for tailoring nanoscale forces via the interplay between collective excitations and structural symmetries, with potential to inspire further exploration of anomalous scaling laws in broader fluctuation-induced phenomena. Our results are expected to be realizable in Rydberg or ultracold atom experiments and accessible via optomechanical measurements or atomic force microscopy in the candidate material platforms.

\bigskip

\begin{acknowledgments}
\textit{Acknowledgments.--}
H.P. and J.-S.W. are supported by the Ministry of Education, Singapore, under the Academic Research Fund Tier 1 (FY2022), grant number  A-8000990-00-00.
The computational work for this article was partially performed on resources of the National Supercomputing Centre, Singapore (\url{https://www.nscc.sg}).
\end{acknowledgments}

\bibliography{main}

\begin{thebibliography}{93}%
\makeatletter
\providecommand \@ifxundefined [1]{%
 \@ifx{#1\undefined}
}%
\providecommand \@ifnum [1]{%
 \ifnum #1\expandafter \@firstoftwo
 \else \expandafter \@secondoftwo
 \fi
}%
\providecommand \@ifx [1]{%
 \ifx #1\expandafter \@firstoftwo
 \else \expandafter \@secondoftwo
 \fi
}%
\providecommand \natexlab [1]{#1}%
\providecommand \enquote  [1]{``#1''}%
\providecommand \bibnamefont  [1]{#1}%
\providecommand \bibfnamefont [1]{#1}%
\providecommand \citenamefont [1]{#1}%
\providecommand \href@noop [0]{\@secondoftwo}%
\providecommand \href [0]{\begingroup \@sanitize@url \@href}%
\providecommand \@href[1]{\@@startlink{#1}\@@href}%
\providecommand \@@href[1]{\endgroup#1\@@endlink}%
\providecommand \@sanitize@url [0]{\catcode `\\12\catcode `\$12\catcode `\&12\catcode `\#12\catcode `\^12\catcode `\_12\catcode `\%12\relax}%
\providecommand \@@startlink[1]{}%
\providecommand \@@endlink[0]{}%
\providecommand \url  [0]{\begingroup\@sanitize@url \@url }%
\providecommand \@url [1]{\endgroup\@href {#1}{\urlprefix }}%
\providecommand \urlprefix  [0]{URL }%
\providecommand \Eprint [0]{\href }%
\providecommand \doibase [0]{https://doi.org/}%
\providecommand \selectlanguage [0]{\@gobble}%
\providecommand \bibinfo  [0]{\@secondoftwo}%
\providecommand \bibfield  [0]{\@secondoftwo}%
\providecommand \translation [1]{[#1]}%
\providecommand \BibitemOpen [0]{}%
\providecommand \bibitemStop [0]{}%
\providecommand \bibitemNoStop [0]{.\EOS\space}%
\providecommand \EOS [0]{\spacefactor3000\relax}%
\providecommand \BibitemShut  [1]{\csname bibitem#1\endcsname}%
\let\auto@bib@innerbib\@empty
\bibitem [{\citenamefont {Parsegian}(2005)}]{parsegian2005van}%
  \BibitemOpen
  \bibfield  {author} {\bibinfo {author} {\bibfnamefont {V.~A.}\ \bibnamefont {Parsegian}},\ }\href@noop {} {\emph {\bibinfo {title} {Van der Waals forces: a handbook for biologists, chemists, engineers, and physicists}}}\ (\bibinfo  {publisher} {Cambridge university press},\ \bibinfo {year} {2005})\BibitemShut {NoStop}%
\bibitem [{\citenamefont {French}\ \emph {et~al.}(2010)\citenamefont {French}, \citenamefont {Parsegian}, \citenamefont {Podgornik}, \citenamefont {Rajter}, \citenamefont {Jagota}, \citenamefont {Luo}, \citenamefont {Asthagiri}, \citenamefont {Chaudhury}, \citenamefont {Chiang}, \citenamefont {Granick}, \citenamefont {Kalinin}, \citenamefont {Kardar}, \citenamefont {Kjellander}, \citenamefont {Langreth}, \citenamefont {Lewis}, \citenamefont {Lustig}, \citenamefont {Wesolowski}, \citenamefont {Wettlaufer}, \citenamefont {Ching}, \citenamefont {Finnis}, \citenamefont {Houlihan}, \citenamefont {von Lilienfeld}, \citenamefont {van Oss},\ and\ \citenamefont {Zemb}}]{French2010}%
  \BibitemOpen
  \bibfield  {author} {\bibinfo {author} {\bibfnamefont {R.~H.}\ \bibnamefont {French}}, \bibinfo {author} {\bibfnamefont {V.~A.}\ \bibnamefont {Parsegian}}, \bibinfo {author} {\bibfnamefont {R.}~\bibnamefont {Podgornik}}, \bibinfo {author} {\bibfnamefont {R.~F.}\ \bibnamefont {Rajter}}, \bibinfo {author} {\bibfnamefont {A.}~\bibnamefont {Jagota}}, \bibinfo {author} {\bibfnamefont {J.}~\bibnamefont {Luo}}, \bibinfo {author} {\bibfnamefont {D.}~\bibnamefont {Asthagiri}}, \bibinfo {author} {\bibfnamefont {M.~K.}\ \bibnamefont {Chaudhury}}, \bibinfo {author} {\bibfnamefont {Y.-m.}\ \bibnamefont {Chiang}}, \bibinfo {author} {\bibfnamefont {S.}~\bibnamefont {Granick}}, \bibinfo {author} {\bibfnamefont {S.}~\bibnamefont {Kalinin}}, \bibinfo {author} {\bibfnamefont {M.}~\bibnamefont {Kardar}}, \bibinfo {author} {\bibfnamefont {R.}~\bibnamefont {Kjellander}}, \bibinfo {author} {\bibfnamefont {D.~C.}\ \bibnamefont {Langreth}}, \bibinfo {author} {\bibfnamefont {J.}~\bibnamefont {Lewis}}, \bibinfo {author}
  {\bibfnamefont {S.}~\bibnamefont {Lustig}}, \bibinfo {author} {\bibfnamefont {D.}~\bibnamefont {Wesolowski}}, \bibinfo {author} {\bibfnamefont {J.~S.}\ \bibnamefont {Wettlaufer}}, \bibinfo {author} {\bibfnamefont {W.-Y.}\ \bibnamefont {Ching}}, \bibinfo {author} {\bibfnamefont {M.}~\bibnamefont {Finnis}}, \bibinfo {author} {\bibfnamefont {F.}~\bibnamefont {Houlihan}}, \bibinfo {author} {\bibfnamefont {O.~A.}\ \bibnamefont {von Lilienfeld}}, \bibinfo {author} {\bibfnamefont {C.~J.}\ \bibnamefont {van Oss}},\ and\ \bibinfo {author} {\bibfnamefont {T.}~\bibnamefont {Zemb}},\ }\bibfield  {title} {\bibinfo {title} {{Long range interactions in nanoscale science}},\ }\href {https://doi.org/10.1103/RevModPhys.82.1887} {\bibfield  {journal} {\bibinfo  {journal} {Rev. Mod. Phys.}\ }\textbf {\bibinfo {volume} {82}},\ \bibinfo {pages} {1887} (\bibinfo {year} {2010})}\BibitemShut {NoStop}%
\bibitem [{\citenamefont {Stone}(2013)}]{Stone2013}%
  \BibitemOpen
  \bibfield  {author} {\bibinfo {author} {\bibfnamefont {A.}~\bibnamefont {Stone}},\ }\href@noop {} {\emph {\bibinfo {title} {{The Theory of Intermolecular Forces}}}},\ \bibinfo {edition} {2nd}\ ed.\ (\bibinfo  {publisher} {Oxford University Press},\ \bibinfo {year} {2013})\BibitemShut {NoStop}%
\bibitem [{\citenamefont {Woods}\ \emph {et~al.}(2016)\citenamefont {Woods}, \citenamefont {Dalvit}, \citenamefont {Tkatchenko}, \citenamefont {Rodriguez-Lopez}, \citenamefont {Rodriguez},\ and\ \citenamefont {Podgornik}}]{Woods2016}%
  \BibitemOpen
  \bibfield  {author} {\bibinfo {author} {\bibfnamefont {L.~M.}\ \bibnamefont {Woods}}, \bibinfo {author} {\bibfnamefont {D.~A.}\ \bibnamefont {Dalvit}}, \bibinfo {author} {\bibfnamefont {A.}~\bibnamefont {Tkatchenko}}, \bibinfo {author} {\bibfnamefont {P.}~\bibnamefont {Rodriguez-Lopez}}, \bibinfo {author} {\bibfnamefont {A.~W.}\ \bibnamefont {Rodriguez}},\ and\ \bibinfo {author} {\bibfnamefont {R.}~\bibnamefont {Podgornik}},\ }\bibfield  {title} {\bibinfo {title} {{Materials perspective on Casimir and van der Waals interactions}},\ }\href {https://doi.org/10.1103/RevModPhys.88.045003} {\bibfield  {journal} {\bibinfo  {journal} {Rev. Mod. Phys.}\ }\textbf {\bibinfo {volume} {88}},\ \bibinfo {pages} {045003} (\bibinfo {year} {2016})}\BibitemShut {NoStop}%
\bibitem [{\citenamefont {Buhmann}(2012)}]{Buhmann2012-2}%
  \BibitemOpen
  \bibfield  {author} {\bibinfo {author} {\bibfnamefont {S.~Y.}\ \bibnamefont {Buhmann}},\ }\href@noop {} {\emph {\bibinfo {title} {{Dispersion Forces II}}}},\ \bibinfo {series} {Springer Tracts in Modern Physics}, Vol.\ \bibinfo {volume} {248}\ (\bibinfo  {publisher} {Springer Berlin Heidelberg},\ \bibinfo {address} {Berlin, Heidelberg},\ \bibinfo {year} {2012})\BibitemShut {NoStop}%
\bibitem [{\citenamefont {Komnik}\ and\ \citenamefont {Egger}(1998)}]{Komnik1998}%
  \BibitemOpen
  \bibfield  {author} {\bibinfo {author} {\bibfnamefont {A.}~\bibnamefont {Komnik}}\ and\ \bibinfo {author} {\bibfnamefont {R.}~\bibnamefont {Egger}},\ }\bibfield  {title} {\bibinfo {title} {{Nonequilibrium transport for crossed Luttinger liquids}},\ }\href {https://doi.org/10.1103/PhysRevLett.80.2881} {\bibfield  {journal} {\bibinfo  {journal} {Phys. Rev. Lett.}\ }\textbf {\bibinfo {volume} {80}},\ \bibinfo {pages} {2881} (\bibinfo {year} {1998})}\BibitemShut {NoStop}%
\bibitem [{\citenamefont {Tse}\ \emph {et~al.}(2007)\citenamefont {Tse}, \citenamefont {Hu},\ and\ \citenamefont {{Das Sarma}}}]{Tse2007}%
  \BibitemOpen
  \bibfield  {author} {\bibinfo {author} {\bibfnamefont {W.-K.}\ \bibnamefont {Tse}}, \bibinfo {author} {\bibfnamefont {B.~Y.-K.}\ \bibnamefont {Hu}},\ and\ \bibinfo {author} {\bibfnamefont {S.}~\bibnamefont {{Das Sarma}}},\ }\bibfield  {title} {\bibinfo {title} {{Theory of Coulomb drag in graphene}},\ }\href {https://doi.org/10.1103/PhysRevB.76.081401} {\bibfield  {journal} {\bibinfo  {journal} {Phys. Rev. B}\ }\textbf {\bibinfo {volume} {76}},\ \bibinfo {pages} {081401} (\bibinfo {year} {2007})}\BibitemShut {NoStop}%
\bibitem [{\citenamefont {Volokitin}\ and\ \citenamefont {Persson}(2011)}]{Volokitin2011}%
  \BibitemOpen
  \bibfield  {author} {\bibinfo {author} {\bibfnamefont {A.~I.}\ \bibnamefont {Volokitin}}\ and\ \bibinfo {author} {\bibfnamefont {B.~N.}\ \bibnamefont {Persson}},\ }\bibfield  {title} {\bibinfo {title} {{Quantum friction}},\ }\href {https://doi.org/10.1103/PhysRevLett.106.094502} {\bibfield  {journal} {\bibinfo  {journal} {Phys. Rev. Lett.}\ }\textbf {\bibinfo {volume} {106}},\ \bibinfo {pages} {094502} (\bibinfo {year} {2011})}\BibitemShut {NoStop}%
\bibitem [{\citenamefont {Basov}\ \emph {et~al.}(2016)\citenamefont {Basov}, \citenamefont {Fogler},\ and\ \citenamefont {{Garc{\'{i}}a De Abajo}}}]{Basov2016}%
  \BibitemOpen
  \bibfield  {author} {\bibinfo {author} {\bibfnamefont {D.~N.}\ \bibnamefont {Basov}}, \bibinfo {author} {\bibfnamefont {M.~M.}\ \bibnamefont {Fogler}},\ and\ \bibinfo {author} {\bibfnamefont {F.~J.}\ \bibnamefont {{Garc{\'{i}}a De Abajo}}},\ }\bibfield  {title} {\bibinfo {title} {{Polaritons in van der Waals materials}},\ }\href {https://www.science.org/doi/10.1126/science.aag1992} {\bibfield  {journal} {\bibinfo  {journal} {Science}\ }\textbf {\bibinfo {volume} {354}} (\bibinfo {year} {2016})}\BibitemShut {NoStop}%
\bibitem [{\citenamefont {Venkataram}\ \emph {et~al.}(2020)\citenamefont {Venkataram}, \citenamefont {Hermann}, \citenamefont {Tkatchenko},\ and\ \citenamefont {Rodriguez}}]{Venkataram2020}%
  \BibitemOpen
  \bibfield  {author} {\bibinfo {author} {\bibfnamefont {P.~S.}\ \bibnamefont {Venkataram}}, \bibinfo {author} {\bibfnamefont {J.}~\bibnamefont {Hermann}}, \bibinfo {author} {\bibfnamefont {A.}~\bibnamefont {Tkatchenko}},\ and\ \bibinfo {author} {\bibfnamefont {A.~W.}\ \bibnamefont {Rodriguez}},\ }\bibfield  {title} {\bibinfo {title} {{Fluctuational electrodynamics in atomic and macroscopic systems: van der Waals interactions and radiative heat transfer}},\ }\href {https://doi.org/10.1103/PhysRevB.102.085403} {\bibfield  {journal} {\bibinfo  {journal} {Phys. Rev. B}\ }\textbf {\bibinfo {volume} {102}},\ \bibinfo {pages} {085403} (\bibinfo {year} {2020})}\BibitemShut {NoStop}%
\bibitem [{\citenamefont {Hauseux}\ \emph {et~al.}(2022)\citenamefont {Hauseux}, \citenamefont {Ambrosetti}, \citenamefont {Bordas},\ and\ \citenamefont {Tkatchenko}}]{Hauseux2022}%
  \BibitemOpen
  \bibfield  {author} {\bibinfo {author} {\bibfnamefont {P.}~\bibnamefont {Hauseux}}, \bibinfo {author} {\bibfnamefont {A.}~\bibnamefont {Ambrosetti}}, \bibinfo {author} {\bibfnamefont {S.~P.}\ \bibnamefont {Bordas}},\ and\ \bibinfo {author} {\bibfnamefont {A.}~\bibnamefont {Tkatchenko}},\ }\bibfield  {title} {\bibinfo {title} {Colossal enhancement of atomic force response in van der waals materials arising from many-body electronic correlations},\ }\href {https://doi.org/10.1103/PhysRevLett.128.106101} {\bibfield  {journal} {\bibinfo  {journal} {Phys. Rev. Lett.}\ }\textbf {\bibinfo {volume} {128}},\ \bibinfo {pages} {106101} (\bibinfo {year} {2022})}\BibitemShut {NoStop}%
\bibitem [{\citenamefont {Chan}\ \emph {et~al.}(2001)\citenamefont {Chan}, \citenamefont {Aksyuk}, \citenamefont {Kleiman}, \citenamefont {Bishop},\ and\ \citenamefont {Capasso}}]{Chan2001}%
  \BibitemOpen
  \bibfield  {author} {\bibinfo {author} {\bibfnamefont {H.~B.}\ \bibnamefont {Chan}}, \bibinfo {author} {\bibfnamefont {V.~A.}\ \bibnamefont {Aksyuk}}, \bibinfo {author} {\bibfnamefont {R.~N.}\ \bibnamefont {Kleiman}}, \bibinfo {author} {\bibfnamefont {D.~J.}\ \bibnamefont {Bishop}},\ and\ \bibinfo {author} {\bibfnamefont {F.}~\bibnamefont {Capasso}},\ }\bibfield  {title} {\bibinfo {title} {{Quantum mechanical actuation of microelectromechanical systems by the Casimir force}},\ }\href {https://doi.org/10.1126/science.1057984} {\bibfield  {journal} {\bibinfo  {journal} {Science}\ }\textbf {\bibinfo {volume} {291}},\ \bibinfo {pages} {1941} (\bibinfo {year} {2001})}\BibitemShut {NoStop}%
\bibitem [{\citenamefont {DelRio}\ \emph {et~al.}(2005)\citenamefont {DelRio}, \citenamefont {de~Boer}, \citenamefont {Knapp}, \citenamefont {{David Reedy}}, \citenamefont {Clews},\ and\ \citenamefont {Dunn}}]{DelRio2005}%
  \BibitemOpen
  \bibfield  {author} {\bibinfo {author} {\bibfnamefont {F.~W.}\ \bibnamefont {DelRio}}, \bibinfo {author} {\bibfnamefont {M.~P.}\ \bibnamefont {de~Boer}}, \bibinfo {author} {\bibfnamefont {J.~A.}\ \bibnamefont {Knapp}}, \bibinfo {author} {\bibfnamefont {E.}~\bibnamefont {{David Reedy}}}, \bibinfo {author} {\bibfnamefont {P.~J.}\ \bibnamefont {Clews}},\ and\ \bibinfo {author} {\bibfnamefont {M.~L.}\ \bibnamefont {Dunn}},\ }\bibfield  {title} {\bibinfo {title} {{The role of van der Waals forces in adhesion of micromachined surfaces}},\ }\href {https://doi.org/10.1038/nmat1431} {\bibfield  {journal} {\bibinfo  {journal} {Nat. Mater.}\ }\textbf {\bibinfo {volume} {4}},\ \bibinfo {pages} {629} (\bibinfo {year} {2005})}\BibitemShut {NoStop}%
\bibitem [{\citenamefont {Bishop}\ \emph {et~al.}(2009)\citenamefont {Bishop}, \citenamefont {Wilmer}, \citenamefont {Soh},\ and\ \citenamefont {Grzybowski}}]{Bishop2009}%
  \BibitemOpen
  \bibfield  {author} {\bibinfo {author} {\bibfnamefont {K.~J.}\ \bibnamefont {Bishop}}, \bibinfo {author} {\bibfnamefont {C.~E.}\ \bibnamefont {Wilmer}}, \bibinfo {author} {\bibfnamefont {S.}~\bibnamefont {Soh}},\ and\ \bibinfo {author} {\bibfnamefont {B.~A.}\ \bibnamefont {Grzybowski}},\ }\bibfield  {title} {\bibinfo {title} {{Nanoscale forces and their uses in self-assembly}},\ }\href {https://doi.org/10.1002/smll.200900358} {\bibfield  {journal} {\bibinfo  {journal} {Small}\ }\textbf {\bibinfo {volume} {5}},\ \bibinfo {pages} {1600} (\bibinfo {year} {2009})}\BibitemShut {NoStop}%
\bibitem [{\citenamefont {Dorrell}\ \emph {et~al.}(2020)\citenamefont {Dorrell}, \citenamefont {Pirie}, \citenamefont {Gardezi}, \citenamefont {Drucker},\ and\ \citenamefont {Hoffman}}]{Dorrell2020}%
  \BibitemOpen
  \bibfield  {author} {\bibinfo {author} {\bibfnamefont {W.}~\bibnamefont {Dorrell}}, \bibinfo {author} {\bibfnamefont {H.}~\bibnamefont {Pirie}}, \bibinfo {author} {\bibfnamefont {S.~M.}\ \bibnamefont {Gardezi}}, \bibinfo {author} {\bibfnamefont {N.~C.}\ \bibnamefont {Drucker}},\ and\ \bibinfo {author} {\bibfnamefont {J.~E.}\ \bibnamefont {Hoffman}},\ }\bibfield  {title} {\bibinfo {title} {{van der Waals metamaterials}},\ }\href {https://doi.org/10.1103/PhysRevB.101.121103} {\bibfield  {journal} {\bibinfo  {journal} {Phys. Rev. B}\ }\textbf {\bibinfo {volume} {101}},\ \bibinfo {pages} {121103} (\bibinfo {year} {2020})}\BibitemShut {NoStop}%
\bibitem [{\citenamefont {Mazzola}\ and\ \citenamefont {Sorella}(2017)}]{Mazzola2017}%
  \BibitemOpen
  \bibfield  {author} {\bibinfo {author} {\bibfnamefont {G.}~\bibnamefont {Mazzola}}\ and\ \bibinfo {author} {\bibfnamefont {S.}~\bibnamefont {Sorella}},\ }\bibfield  {title} {\bibinfo {title} {{Accelerating ab initio molecular dynamics and probing the weak dispersive forces in dense liquid hydrogen}},\ }\href {https://doi.org/10.1103/PhysRevLett.118.015703} {\bibfield  {journal} {\bibinfo  {journal} {Phys. Rev. Lett.}\ }\textbf {\bibinfo {volume} {118}},\ \bibinfo {pages} {015703} (\bibinfo {year} {2017})}\BibitemShut {NoStop}%
\bibitem [{\citenamefont {Nerngchamnong}\ \emph {et~al.}(2013)\citenamefont {Nerngchamnong}, \citenamefont {Yuan}, \citenamefont {Qi}, \citenamefont {Li}, \citenamefont {Thompson},\ and\ \citenamefont {Nijhuis}}]{Nerngchamnong2013}%
  \BibitemOpen
  \bibfield  {author} {\bibinfo {author} {\bibfnamefont {N.}~\bibnamefont {Nerngchamnong}}, \bibinfo {author} {\bibfnamefont {L.}~\bibnamefont {Yuan}}, \bibinfo {author} {\bibfnamefont {D.-C.}\ \bibnamefont {Qi}}, \bibinfo {author} {\bibfnamefont {J.}~\bibnamefont {Li}}, \bibinfo {author} {\bibfnamefont {D.}~\bibnamefont {Thompson}},\ and\ \bibinfo {author} {\bibfnamefont {C.~A.}\ \bibnamefont {Nijhuis}},\ }\bibfield  {title} {\bibinfo {title} {{The role of van der Waals forces in the performance of molecular diodes}},\ }\href {https://doi.org/10.1038/nnano.2012.238} {\bibfield  {journal} {\bibinfo  {journal} {Nat. Nanotechnol.}\ }\textbf {\bibinfo {volume} {8}},\ \bibinfo {pages} {113} (\bibinfo {year} {2013})}\BibitemShut {NoStop}%
\bibitem [{\citenamefont {Dobson}\ \emph {et~al.}(2001)\citenamefont {Dobson}, \citenamefont {McLennan}, \citenamefont {Rubio}, \citenamefont {Wang}, \citenamefont {Gould}, \citenamefont {Le},\ and\ \citenamefont {Dinte}}]{Dobson2001}%
  \BibitemOpen
  \bibfield  {author} {\bibinfo {author} {\bibfnamefont {J.~F.}\ \bibnamefont {Dobson}}, \bibinfo {author} {\bibfnamefont {K.}~\bibnamefont {McLennan}}, \bibinfo {author} {\bibfnamefont {A.}~\bibnamefont {Rubio}}, \bibinfo {author} {\bibfnamefont {J.}~\bibnamefont {Wang}}, \bibinfo {author} {\bibfnamefont {T.}~\bibnamefont {Gould}}, \bibinfo {author} {\bibfnamefont {H.~M.}\ \bibnamefont {Le}},\ and\ \bibinfo {author} {\bibfnamefont {B.~P.}\ \bibnamefont {Dinte}},\ }\bibfield  {title} {\bibinfo {title} {{Prediction of dispersion forces: Is there a problem?}},\ }\href {https://doi.org/10.1071/CH01052} {\bibfield  {journal} {\bibinfo  {journal} {Aust. J. Chem.}\ }\textbf {\bibinfo {volume} {54}},\ \bibinfo {pages} {513} (\bibinfo {year} {2001})}\BibitemShut {NoStop}%
\bibitem [{\citenamefont {Dobson}\ \emph {et~al.}(2006)\citenamefont {Dobson}, \citenamefont {White},\ and\ \citenamefont {Rubio}}]{Dobson2006}%
  \BibitemOpen
  \bibfield  {author} {\bibinfo {author} {\bibfnamefont {J.~F.}\ \bibnamefont {Dobson}}, \bibinfo {author} {\bibfnamefont {A.}~\bibnamefont {White}},\ and\ \bibinfo {author} {\bibfnamefont {A.}~\bibnamefont {Rubio}},\ }\bibfield  {title} {\bibinfo {title} {{Asymptotics of the dispersion interaction: Analytic benchmarks for van der Waals energy functionals}},\ }\href {https://doi.org/10.1103/PhysRevLett.96.073201} {\bibfield  {journal} {\bibinfo  {journal} {Phys. Rev. Lett.}\ }\textbf {\bibinfo {volume} {96}},\ \bibinfo {pages} {073201} (\bibinfo {year} {2006})}\BibitemShut {NoStop}%
\bibitem [{\citenamefont {Tkatchenko}\ and\ \citenamefont {Scheffler}(2009)}]{Tkatchenko2009}%
  \BibitemOpen
  \bibfield  {author} {\bibinfo {author} {\bibfnamefont {A.}~\bibnamefont {Tkatchenko}}\ and\ \bibinfo {author} {\bibfnamefont {M.}~\bibnamefont {Scheffler}},\ }\bibfield  {title} {\bibinfo {title} {{Accurate molecular van der Waals interactions from ground-state electron density and free-atom reference data}},\ }\href {https://doi.org/10.1103/PhysRevLett.102.073005} {\bibfield  {journal} {\bibinfo  {journal} {Phys. Rev. Lett.}\ }\textbf {\bibinfo {volume} {102}},\ \bibinfo {pages} {073005} (\bibinfo {year} {2009})}\BibitemShut {NoStop}%
\bibitem [{\citenamefont {Cole}\ \emph {et~al.}(2009)\citenamefont {Cole}, \citenamefont {Velegol}, \citenamefont {Kim},\ and\ \citenamefont {Lucas}}]{Cole2009}%
  \BibitemOpen
  \bibfield  {author} {\bibinfo {author} {\bibfnamefont {M.~W.}\ \bibnamefont {Cole}}, \bibinfo {author} {\bibfnamefont {D.}~\bibnamefont {Velegol}}, \bibinfo {author} {\bibfnamefont {H.-Y.}\ \bibnamefont {Kim}},\ and\ \bibinfo {author} {\bibfnamefont {A.~A.}\ \bibnamefont {Lucas}},\ }\bibfield  {title} {\bibinfo {title} {{Nanoscale van der Waals interactions}},\ }\href {https://doi.org/10.1080/08927020902929794} {\bibfield  {journal} {\bibinfo  {journal} {Mol. Simul.}\ }\textbf {\bibinfo {volume} {35}},\ \bibinfo {pages} {849} (\bibinfo {year} {2009})}\BibitemShut {NoStop}%
\bibitem [{\citenamefont {Leb{\`{e}}gue}\ \emph {et~al.}(2010)\citenamefont {Leb{\`{e}}gue}, \citenamefont {Harl}, \citenamefont {Gould}, \citenamefont {{\'{A}}ngy{\'{a}}n}, \citenamefont {Kresse},\ and\ \citenamefont {Dobson}}]{Lebegue2010}%
  \BibitemOpen
  \bibfield  {author} {\bibinfo {author} {\bibfnamefont {S.}~\bibnamefont {Leb{\`{e}}gue}}, \bibinfo {author} {\bibfnamefont {J.}~\bibnamefont {Harl}}, \bibinfo {author} {\bibfnamefont {T.}~\bibnamefont {Gould}}, \bibinfo {author} {\bibfnamefont {J.~G.}\ \bibnamefont {{\'{A}}ngy{\'{a}}n}}, \bibinfo {author} {\bibfnamefont {G.}~\bibnamefont {Kresse}},\ and\ \bibinfo {author} {\bibfnamefont {J.~F.}\ \bibnamefont {Dobson}},\ }\bibfield  {title} {\bibinfo {title} {{Cohesive properties and asymptotics of the dispersion interaction in graphite by the random phase approximation}},\ }\href {https://doi.org/10.1103/PhysRevLett.105.196401} {\bibfield  {journal} {\bibinfo  {journal} {Phys. Rev. Lett.}\ }\textbf {\bibinfo {volume} {105}},\ \bibinfo {pages} {196401} (\bibinfo {year} {2010})}\BibitemShut {NoStop}%
\bibitem [{\citenamefont {Sarabadani}\ \emph {et~al.}(2011)\citenamefont {Sarabadani}, \citenamefont {Naji}, \citenamefont {Asgari},\ and\ \citenamefont {Podgornik}}]{Sarabadani2011}%
  \BibitemOpen
  \bibfield  {author} {\bibinfo {author} {\bibfnamefont {J.}~\bibnamefont {Sarabadani}}, \bibinfo {author} {\bibfnamefont {A.}~\bibnamefont {Naji}}, \bibinfo {author} {\bibfnamefont {R.}~\bibnamefont {Asgari}},\ and\ \bibinfo {author} {\bibfnamefont {R.}~\bibnamefont {Podgornik}},\ }\bibfield  {title} {\bibinfo {title} {{Many-body effects in the van der Waals-Casimir interaction between graphene layers}},\ }\href {https://doi.org/10.1103/PhysRevB.84.155407} {\bibfield  {journal} {\bibinfo  {journal} {Phys. Rev. B}\ }\textbf {\bibinfo {volume} {84}},\ \bibinfo {pages} {155407} (\bibinfo {year} {2011})}\BibitemShut {NoStop}%
\bibitem [{\citenamefont {Tkatchenko}\ \emph {et~al.}(2012)\citenamefont {Tkatchenko}, \citenamefont {Distasio}, \citenamefont {Car},\ and\ \citenamefont {Scheffler}}]{Tkatchenko2012}%
  \BibitemOpen
  \bibfield  {author} {\bibinfo {author} {\bibfnamefont {A.}~\bibnamefont {Tkatchenko}}, \bibinfo {author} {\bibfnamefont {R.~A.}\ \bibnamefont {Distasio}}, \bibinfo {author} {\bibfnamefont {R.}~\bibnamefont {Car}},\ and\ \bibinfo {author} {\bibfnamefont {M.}~\bibnamefont {Scheffler}},\ }\bibfield  {title} {\bibinfo {title} {{Accurate and efficient method for many-body van der Waals interactions}},\ }\href {https://doi.org/10.1103/PhysRevLett.108.236402} {\bibfield  {journal} {\bibinfo  {journal} {Phys. Rev. Lett.}\ }\textbf {\bibinfo {volume} {108}},\ \bibinfo {pages} {236402} (\bibinfo {year} {2012})}\BibitemShut {NoStop}%
\bibitem [{\citenamefont {Noruzifar}\ \emph {et~al.}(2012)\citenamefont {Noruzifar}, \citenamefont {Emig}, \citenamefont {Mohideen},\ and\ \citenamefont {Zandi}}]{Noruzifar2012}%
  \BibitemOpen
  \bibfield  {author} {\bibinfo {author} {\bibfnamefont {E.}~\bibnamefont {Noruzifar}}, \bibinfo {author} {\bibfnamefont {T.}~\bibnamefont {Emig}}, \bibinfo {author} {\bibfnamefont {U.}~\bibnamefont {Mohideen}},\ and\ \bibinfo {author} {\bibfnamefont {R.}~\bibnamefont {Zandi}},\ }\bibfield  {title} {\bibinfo {title} {{Collective charge fluctuations and Casimir interactions for quasi-one-dimensional metals}},\ }\href {https://doi.org/10.1103/PhysRevB.86.115449} {\bibfield  {journal} {\bibinfo  {journal} {Phys. Rev. B}\ }\textbf {\bibinfo {volume} {86}},\ \bibinfo {pages} {115449} (\bibinfo {year} {2012})}\BibitemShut {NoStop}%
\bibitem [{\citenamefont {Dobson}\ \emph {et~al.}(2014)\citenamefont {Dobson}, \citenamefont {Gould},\ and\ \citenamefont {Vignale}}]{Dobson2014How}%
  \BibitemOpen
  \bibfield  {author} {\bibinfo {author} {\bibfnamefont {J.~F.}\ \bibnamefont {Dobson}}, \bibinfo {author} {\bibfnamefont {T.}~\bibnamefont {Gould}},\ and\ \bibinfo {author} {\bibfnamefont {G.}~\bibnamefont {Vignale}},\ }\bibfield  {title} {\bibinfo {title} {{How many-body effects modify the van der waals interaction between graphene sheets}},\ }\href {https://doi.org/10.1103/PhysRevX.4.021040} {\bibfield  {journal} {\bibinfo  {journal} {Phys. Rev. X}\ }\textbf {\bibinfo {volume} {4}},\ \bibinfo {pages} {021040} (\bibinfo {year} {2014})}\BibitemShut {NoStop}%
\bibitem [{\citenamefont {Dobson}(2014)}]{Dobson2014}%
  \BibitemOpen
  \bibfield  {author} {\bibinfo {author} {\bibfnamefont {J.~F.}\ \bibnamefont {Dobson}},\ }\bibfield  {title} {\bibinfo {title} {{Beyond pairwise additivity in London dispersion interactions}},\ }\href {https://doi.org/10.1002/qua.24635} {\bibfield  {journal} {\bibinfo  {journal} {Int. J. Quantum Chem.}\ }\textbf {\bibinfo {volume} {114}},\ \bibinfo {pages} {1157} (\bibinfo {year} {2014})}\BibitemShut {NoStop}%
\bibitem [{\citenamefont {Wagner}\ \emph {et~al.}(2014)\citenamefont {Wagner}, \citenamefont {Fournier}, \citenamefont {Ruiz}, \citenamefont {Li}, \citenamefont {M{\"{u}}llen}, \citenamefont {Rohlfing}, \citenamefont {Tkatchenko}, \citenamefont {Temirov},\ and\ \citenamefont {Tautz}}]{Wagner2014}%
  \BibitemOpen
  \bibfield  {author} {\bibinfo {author} {\bibfnamefont {C.}~\bibnamefont {Wagner}}, \bibinfo {author} {\bibfnamefont {N.}~\bibnamefont {Fournier}}, \bibinfo {author} {\bibfnamefont {V.~G.}\ \bibnamefont {Ruiz}}, \bibinfo {author} {\bibfnamefont {C.}~\bibnamefont {Li}}, \bibinfo {author} {\bibfnamefont {K.}~\bibnamefont {M{\"{u}}llen}}, \bibinfo {author} {\bibfnamefont {M.}~\bibnamefont {Rohlfing}}, \bibinfo {author} {\bibfnamefont {A.}~\bibnamefont {Tkatchenko}}, \bibinfo {author} {\bibfnamefont {R.}~\bibnamefont {Temirov}},\ and\ \bibinfo {author} {\bibfnamefont {F.~S.}\ \bibnamefont {Tautz}},\ }\bibfield  {title} {\bibinfo {title} {{Non-additivity of molecule-surface van der Waals potentials from force measurements}},\ }\href {https://doi.org/10.1038/ncomms6568} {\bibfield  {journal} {\bibinfo  {journal} {Nat. Commun.}\ }\textbf {\bibinfo {volume} {5}},\ \bibinfo {pages} {5568} (\bibinfo {year} {2014})}\BibitemShut {NoStop}%
\bibitem [{\citenamefont {Reilly}\ and\ \citenamefont {Tkatchenko}(2015)}]{Reilly2015}%
  \BibitemOpen
  \bibfield  {author} {\bibinfo {author} {\bibfnamefont {A.~M.}\ \bibnamefont {Reilly}}\ and\ \bibinfo {author} {\bibfnamefont {A.}~\bibnamefont {Tkatchenko}},\ }\bibfield  {title} {\bibinfo {title} {{Van der Waals dispersion interactions in molecular materials: Beyond pairwise additivity}},\ }\href {https://doi.org/10.1039/c5sc00410a} {\bibfield  {journal} {\bibinfo  {journal} {Chem. Sci.}\ }\textbf {\bibinfo {volume} {6}},\ \bibinfo {pages} {3289} (\bibinfo {year} {2015})}\BibitemShut {NoStop}%
\bibitem [{\citenamefont {Ambrosetti}\ \emph {et~al.}(2016)\citenamefont {Ambrosetti}, \citenamefont {Ferri}, \citenamefont {DiStasio},\ and\ \citenamefont {Tkatchenko}}]{Ambrosetti2016}%
  \BibitemOpen
  \bibfield  {author} {\bibinfo {author} {\bibfnamefont {A.}~\bibnamefont {Ambrosetti}}, \bibinfo {author} {\bibfnamefont {N.}~\bibnamefont {Ferri}}, \bibinfo {author} {\bibfnamefont {R.~A.}\ \bibnamefont {DiStasio}},\ and\ \bibinfo {author} {\bibfnamefont {A.}~\bibnamefont {Tkatchenko}},\ }\bibfield  {title} {\bibinfo {title} {{Wavelike charge density fluctuations and van der Waals interactions at the nanoscale}},\ }\href {https://doi.org/10.1126/science.aae0509} {\bibfield  {journal} {\bibinfo  {journal} {Science}\ }\textbf {\bibinfo {volume} {351}},\ \bibinfo {pages} {1171} (\bibinfo {year} {2016})}\BibitemShut {NoStop}%
\bibitem [{\citenamefont {Hermann}\ and\ \citenamefont {Tkatchenko}(2020)}]{Hermann2020}%
  \BibitemOpen
  \bibfield  {author} {\bibinfo {author} {\bibfnamefont {J.}~\bibnamefont {Hermann}}\ and\ \bibinfo {author} {\bibfnamefont {A.}~\bibnamefont {Tkatchenko}},\ }\bibfield  {title} {\bibinfo {title} {{Density functional model for van der Waals interactions: Unifying many-body atomic approaches with nonlocal functionals}},\ }\href {https://doi.org/10.1103/PhysRevLett.124.146401} {\bibfield  {journal} {\bibinfo  {journal} {Phys. Rev. Lett.}\ }\textbf {\bibinfo {volume} {124}},\ \bibinfo {pages} {146401} (\bibinfo {year} {2020})}\BibitemShut {NoStop}%
\bibitem [{\citenamefont {Gould}\ \emph {et~al.}(2009)\citenamefont {Gould}, \citenamefont {Gray},\ and\ \citenamefont {Dobson}}]{Gould2009}%
  \BibitemOpen
  \bibfield  {author} {\bibinfo {author} {\bibfnamefont {T.}~\bibnamefont {Gould}}, \bibinfo {author} {\bibfnamefont {E.}~\bibnamefont {Gray}},\ and\ \bibinfo {author} {\bibfnamefont {J.~F.}\ \bibnamefont {Dobson}},\ }\bibfield  {title} {\bibinfo {title} {{van der Waals dispersion power laws for cleavage, exfoliation, and stretching in multiscale, layered systems}},\ }\href {https://doi.org/10.1103/PhysRevB.79.113402} {\bibfield  {journal} {\bibinfo  {journal} {Phys. Rev. B}\ }\textbf {\bibinfo {volume} {79}},\ \bibinfo {pages} {113402} (\bibinfo {year} {2009})}\BibitemShut {NoStop}%
\bibitem [{\citenamefont {Gobre}\ and\ \citenamefont {Tkatchenko}(2013)}]{Gobre2013}%
  \BibitemOpen
  \bibfield  {author} {\bibinfo {author} {\bibfnamefont {V.~V.}\ \bibnamefont {Gobre}}\ and\ \bibinfo {author} {\bibfnamefont {A.}~\bibnamefont {Tkatchenko}},\ }\bibfield  {title} {\bibinfo {title} {{Scaling laws for van der Waals interactions in nanostructured materials}},\ }\href {https://doi.org/10.1038/ncomms3341} {\bibfield  {journal} {\bibinfo  {journal} {Nat. Commun.}\ }\textbf {\bibinfo {volume} {4}},\ \bibinfo {pages} {2341} (\bibinfo {year} {2013})}\BibitemShut {NoStop}%
\bibitem [{\citenamefont {Stedman}\ \emph {et~al.}(2014)\citenamefont {Stedman}, \citenamefont {Drosdoff},\ and\ \citenamefont {Woods}}]{Stedman2014}%
  \BibitemOpen
  \bibfield  {author} {\bibinfo {author} {\bibfnamefont {T.}~\bibnamefont {Stedman}}, \bibinfo {author} {\bibfnamefont {D.}~\bibnamefont {Drosdoff}},\ and\ \bibinfo {author} {\bibfnamefont {L.~M.}\ \bibnamefont {Woods}},\ }\bibfield  {title} {\bibinfo {title} {{van der Waals interactions between nanostructures: Some analytic results from series expansions}},\ }\href {https://doi.org/10.1103/PhysRevA.89.012509} {\bibfield  {journal} {\bibinfo  {journal} {Phys. Rev. A}\ }\textbf {\bibinfo {volume} {89}},\ \bibinfo {pages} {012509} (\bibinfo {year} {2014})}\BibitemShut {NoStop}%
\bibitem [{\citenamefont {Makhnovets}\ and\ \citenamefont {Kolezhuk}(2017)}]{Makhnovets2017}%
  \BibitemOpen
  \bibfield  {author} {\bibinfo {author} {\bibfnamefont {K.~A.}\ \bibnamefont {Makhnovets}}\ and\ \bibinfo {author} {\bibfnamefont {A.~K.}\ \bibnamefont {Kolezhuk}},\ }\bibfield  {title} {\bibinfo {title} {{Finite-size nanowire at a surface: Unconventional power laws of the van der Waals interaction}},\ }\href {https://doi.org/10.1103/PhysRevB.96.125427} {\bibfield  {journal} {\bibinfo  {journal} {Phys. Rev. B}\ }\textbf {\bibinfo {volume} {96}},\ \bibinfo {pages} {125427} (\bibinfo {year} {2017})}\BibitemShut {NoStop}%
\bibitem [{\citenamefont {Yang}\ \emph {et~al.}(2019)\citenamefont {Yang}, \citenamefont {Lao},\ and\ \citenamefont {Distasio}}]{Yang2019}%
  \BibitemOpen
  \bibfield  {author} {\bibinfo {author} {\bibfnamefont {Y.}~\bibnamefont {Yang}}, \bibinfo {author} {\bibfnamefont {K.~U.}\ \bibnamefont {Lao}},\ and\ \bibinfo {author} {\bibfnamefont {R.~A.}\ \bibnamefont {Distasio}},\ }\bibfield  {title} {\bibinfo {title} {Influence of pore size on the van der waals interaction in two-dimensional molecules and materials},\ }\href {https://doi.org/10.1103/PhysRevLett.122.026001} {\bibfield  {journal} {\bibinfo  {journal} {Phys. Rev. Lett.}\ }\textbf {\bibinfo {volume} {122}},\ \bibinfo {pages} {026001} (\bibinfo {year} {2019})}\BibitemShut {NoStop}%
\bibitem [{\citenamefont {Kou}\ \emph {et~al.}(2023)\citenamefont {Kou}, \citenamefont {Chen}, \citenamefont {Jiang}, \citenamefont {Guo},\ and\ \citenamefont {Liu}}]{Kou2023}%
  \BibitemOpen
  \bibfield  {author} {\bibinfo {author} {\bibfnamefont {Z.}~\bibnamefont {Kou}}, \bibinfo {author} {\bibfnamefont {F.}~\bibnamefont {Chen}}, \bibinfo {author} {\bibfnamefont {Z.}~\bibnamefont {Jiang}}, \bibinfo {author} {\bibfnamefont {W.}~\bibnamefont {Guo}},\ and\ \bibinfo {author} {\bibfnamefont {X.}~\bibnamefont {Liu}},\ }\bibfield  {title} {\bibinfo {title} {{Interedge van der Waals interaction between two-dimensional materials}},\ }\href {https://doi.org/10.1103/PhysRevB.108.115420} {\bibfield  {journal} {\bibinfo  {journal} {Phys. Rev. B}\ }\textbf {\bibinfo {volume} {108}},\ \bibinfo {pages} {115420} (\bibinfo {year} {2023})}\BibitemShut {NoStop}%
\bibitem [{\citenamefont {Le}\ \emph {et~al.}(2024{\natexlab{a}})\citenamefont {Le}, \citenamefont {Rodriguez-Lopez},\ and\ \citenamefont {Woods}}]{Le2024}%
  \BibitemOpen
  \bibfield  {author} {\bibinfo {author} {\bibfnamefont {D.-N.}\ \bibnamefont {Le}}, \bibinfo {author} {\bibfnamefont {P.}~\bibnamefont {Rodriguez-Lopez}},\ and\ \bibinfo {author} {\bibfnamefont {L.~M.}\ \bibnamefont {Woods}},\ }\bibfield  {title} {\bibinfo {title} {{Nonlinear effects in manybody van der Waals interactions}},\ }\href {https://doi.org/10.1103/PhysRevResearch.6.013289} {\bibfield  {journal} {\bibinfo  {journal} {Phys. Rev. Res.}\ }\textbf {\bibinfo {volume} {6}},\ \bibinfo {pages} {013289} (\bibinfo {year} {2024}{\natexlab{a}})}\BibitemShut {NoStop}%
\bibitem [{\citenamefont {Filleter}\ \emph {et~al.}(2009)\citenamefont {Filleter}, \citenamefont {McChesney}, \citenamefont {Bostwick}, \citenamefont {Rotenberg}, \citenamefont {Emtsev}, \citenamefont {Seyller}, \citenamefont {Horn},\ and\ \citenamefont {Bennewitz}}]{Filleter2009}%
  \BibitemOpen
  \bibfield  {author} {\bibinfo {author} {\bibfnamefont {T.}~\bibnamefont {Filleter}}, \bibinfo {author} {\bibfnamefont {J.~L.}\ \bibnamefont {McChesney}}, \bibinfo {author} {\bibfnamefont {A.}~\bibnamefont {Bostwick}}, \bibinfo {author} {\bibfnamefont {E.}~\bibnamefont {Rotenberg}}, \bibinfo {author} {\bibfnamefont {K.~V.}\ \bibnamefont {Emtsev}}, \bibinfo {author} {\bibfnamefont {T.}~\bibnamefont {Seyller}}, \bibinfo {author} {\bibfnamefont {K.}~\bibnamefont {Horn}},\ and\ \bibinfo {author} {\bibfnamefont {R.}~\bibnamefont {Bennewitz}},\ }\bibfield  {title} {\bibinfo {title} {{Friction and dissipation in epitaxial graphene films}},\ }\href {https://doi.org/10.1103/PhysRevLett.102.086102} {\bibfield  {journal} {\bibinfo  {journal} {Phys. Rev. Lett.}\ }\textbf {\bibinfo {volume} {102}},\ \bibinfo {pages} {086102} (\bibinfo {year} {2009})}\BibitemShut {NoStop}%
\bibitem [{\citenamefont {Lee}\ \emph {et~al.}(2015)\citenamefont {Lee}, \citenamefont {Avsar}, \citenamefont {Jung}, \citenamefont {Tan}, \citenamefont {Watanabe}, \citenamefont {Taniguchi}, \citenamefont {Natarajan}, \citenamefont {Eda}, \citenamefont {Adam}, \citenamefont {{Castro Neto}},\ and\ \citenamefont {{\"{O}}zyilmaz}}]{Lee2015}%
  \BibitemOpen
  \bibfield  {author} {\bibinfo {author} {\bibfnamefont {J.~H.}\ \bibnamefont {Lee}}, \bibinfo {author} {\bibfnamefont {A.}~\bibnamefont {Avsar}}, \bibinfo {author} {\bibfnamefont {J.}~\bibnamefont {Jung}}, \bibinfo {author} {\bibfnamefont {J.~Y.}\ \bibnamefont {Tan}}, \bibinfo {author} {\bibfnamefont {K.}~\bibnamefont {Watanabe}}, \bibinfo {author} {\bibfnamefont {T.}~\bibnamefont {Taniguchi}}, \bibinfo {author} {\bibfnamefont {S.}~\bibnamefont {Natarajan}}, \bibinfo {author} {\bibfnamefont {G.}~\bibnamefont {Eda}}, \bibinfo {author} {\bibfnamefont {S.}~\bibnamefont {Adam}}, \bibinfo {author} {\bibfnamefont {A.~H.}\ \bibnamefont {{Castro Neto}}},\ and\ \bibinfo {author} {\bibfnamefont {B.}~\bibnamefont {{\"{O}}zyilmaz}},\ }\bibfield  {title} {\bibinfo {title} {{Van der Waals force: A dominant factor for reactivity of graphene}},\ }\href {https://doi.org/10.1021/nl5036012} {\bibfield  {journal} {\bibinfo  {journal} {Nano Lett.}\ }\textbf {\bibinfo {volume} {15}},\ \bibinfo {pages} {319} (\bibinfo {year}
  {2015})}\BibitemShut {NoStop}%
\bibitem [{\citenamefont {Kleshchonok}\ and\ \citenamefont {Tkatchenko}(2018)}]{Kleshchonok2018}%
  \BibitemOpen
  \bibfield  {author} {\bibinfo {author} {\bibfnamefont {A.}~\bibnamefont {Kleshchonok}}\ and\ \bibinfo {author} {\bibfnamefont {A.}~\bibnamefont {Tkatchenko}},\ }\bibfield  {title} {\bibinfo {title} {{Tailoring van der Waals dispersion interactions with external electric charges}},\ }\href {https://doi.org/10.1038/s41467-018-05407-x} {\bibfield  {journal} {\bibinfo  {journal} {Nat. Commun.}\ }\textbf {\bibinfo {volume} {9}},\ \bibinfo {pages} {3017} (\bibinfo {year} {2018})}\BibitemShut {NoStop}%
\bibitem [{\citenamefont {Calbi}\ \emph {et~al.}(2003)\citenamefont {Calbi}, \citenamefont {Gatica}, \citenamefont {Velegol},\ and\ \citenamefont {Cole}}]{Calbi2003}%
  \BibitemOpen
  \bibfield  {author} {\bibinfo {author} {\bibfnamefont {M.~M.}\ \bibnamefont {Calbi}}, \bibinfo {author} {\bibfnamefont {S.~M.}\ \bibnamefont {Gatica}}, \bibinfo {author} {\bibfnamefont {D.}~\bibnamefont {Velegol}},\ and\ \bibinfo {author} {\bibfnamefont {M.~W.}\ \bibnamefont {Cole}},\ }\bibfield  {title} {\bibinfo {title} {{Retarded and nonretarded van der Waals interactions between a cluster and a second cluster or a conducting surface}},\ }\href {https://doi.org/10.1103/PhysRevA.67.033201} {\bibfield  {journal} {\bibinfo  {journal} {Phys. Rev. A}\ }\textbf {\bibinfo {volume} {67}},\ \bibinfo {pages} {5} (\bibinfo {year} {2003})}\BibitemShut {NoStop}%
\bibitem [{\citenamefont {Cho}\ and\ \citenamefont {Silbey}(1996)}]{Cho1996}%
  \BibitemOpen
  \bibfield  {author} {\bibinfo {author} {\bibfnamefont {M.}~\bibnamefont {Cho}}\ and\ \bibinfo {author} {\bibfnamefont {R.~J.}\ \bibnamefont {Silbey}},\ }\bibfield  {title} {\bibinfo {title} {{Suppression and enhancement of van der Waals interactions}},\ }\href {https://doi.org/10.1063/1.471562} {\bibfield  {journal} {\bibinfo  {journal} {J. Chem. Phys.}\ }\textbf {\bibinfo {volume} {104}},\ \bibinfo {pages} {8730} (\bibinfo {year} {1996})}\BibitemShut {NoStop}%
\bibitem [{\citenamefont {Shahmoon}\ and\ \citenamefont {Kurizki}(2013)}]{Shahmoon2013}%
  \BibitemOpen
  \bibfield  {author} {\bibinfo {author} {\bibfnamefont {E.}~\bibnamefont {Shahmoon}}\ and\ \bibinfo {author} {\bibfnamefont {G.}~\bibnamefont {Kurizki}},\ }\bibfield  {title} {\bibinfo {title} {{Dispersion forces inside metallic waveguides}},\ }\href {https://doi.org/10.1103/PhysRevA.87.062105} {\bibfield  {journal} {\bibinfo  {journal} {Phys. Rev. A}\ }\textbf {\bibinfo {volume} {87}},\ \bibinfo {pages} {062105} (\bibinfo {year} {2013})}\BibitemShut {NoStop}%
\bibitem [{\citenamefont {Shahmoon}\ \emph {et~al.}(2014)\citenamefont {Shahmoon}, \citenamefont {Mazets},\ and\ \citenamefont {Kurizki}}]{Shahmoon2014}%
  \BibitemOpen
  \bibfield  {author} {\bibinfo {author} {\bibfnamefont {E.}~\bibnamefont {Shahmoon}}, \bibinfo {author} {\bibfnamefont {I.}~\bibnamefont {Mazets}},\ and\ \bibinfo {author} {\bibfnamefont {G.}~\bibnamefont {Kurizki}},\ }\bibfield  {title} {\bibinfo {title} {{Giant vacuum forces via transmission lines}},\ }\href {https://doi.org/10.1073/pnas.1401346111} {\bibfield  {journal} {\bibinfo  {journal} {Proc. Natl. Acad. Sci.}\ }\textbf {\bibinfo {volume} {111}},\ \bibinfo {pages} {10485} (\bibinfo {year} {2014})}\BibitemShut {NoStop}%
\bibitem [{\citenamefont {B{\'{e}}guin}\ \emph {et~al.}(2013)\citenamefont {B{\'{e}}guin}, \citenamefont {Vernier}, \citenamefont {Chicireanu}, \citenamefont {Lahaye},\ and\ \citenamefont {Browaeys}}]{Beguin2013}%
  \BibitemOpen
  \bibfield  {author} {\bibinfo {author} {\bibfnamefont {L.}~\bibnamefont {B{\'{e}}guin}}, \bibinfo {author} {\bibfnamefont {A.}~\bibnamefont {Vernier}}, \bibinfo {author} {\bibfnamefont {R.}~\bibnamefont {Chicireanu}}, \bibinfo {author} {\bibfnamefont {T.}~\bibnamefont {Lahaye}},\ and\ \bibinfo {author} {\bibfnamefont {A.}~\bibnamefont {Browaeys}},\ }\bibfield  {title} {\bibinfo {title} {{Direct measurement of the van der waals interaction between two rydberg atoms}},\ }\href {https://doi.org/10.1103/PhysRevLett.110.263201} {\bibfield  {journal} {\bibinfo  {journal} {Phys. Rev. Lett.}\ }\textbf {\bibinfo {volume} {110}},\ \bibinfo {pages} {263201} (\bibinfo {year} {2013})}\BibitemShut {NoStop}%
\bibitem [{\citenamefont {Ambrosetti}\ \emph {et~al.}(2022)\citenamefont {Ambrosetti}, \citenamefont {Umari}, \citenamefont {Silvestrelli}, \citenamefont {Elliott},\ and\ \citenamefont {Tkatchenko}}]{Ambrosetti2022}%
  \BibitemOpen
  \bibfield  {author} {\bibinfo {author} {\bibfnamefont {A.}~\bibnamefont {Ambrosetti}}, \bibinfo {author} {\bibfnamefont {P.}~\bibnamefont {Umari}}, \bibinfo {author} {\bibfnamefont {P.~L.}\ \bibnamefont {Silvestrelli}}, \bibinfo {author} {\bibfnamefont {J.}~\bibnamefont {Elliott}},\ and\ \bibinfo {author} {\bibfnamefont {A.}~\bibnamefont {Tkatchenko}},\ }\bibfield  {title} {\bibinfo {title} {{Optical van-der-Waals forces in molecules: from electronic Bethe-Salpeter calculations to the many-body dispersion model}},\ }\href {https://doi.org/10.1038/s41467-022-28461-y} {\bibfield  {journal} {\bibinfo  {journal} {Nat. Commun.}\ }\textbf {\bibinfo {volume} {13}},\ \bibinfo {pages} {813} (\bibinfo {year} {2022})}\BibitemShut {NoStop}%
\bibitem [{\citenamefont {Tasci}\ \emph {et~al.}(2025)\citenamefont {Tasci}, \citenamefont {Cunha},\ and\ \citenamefont {Flick}}]{Tasci2025}%
  \BibitemOpen
  \bibfield  {author} {\bibinfo {author} {\bibfnamefont {C.}~\bibnamefont {Tasci}}, \bibinfo {author} {\bibfnamefont {L.~A.}\ \bibnamefont {Cunha}},\ and\ \bibinfo {author} {\bibfnamefont {J.}~\bibnamefont {Flick}},\ }\bibfield  {title} {\bibinfo {title} {{Photon many-body dispersion: Exchange-correlation functional for strongly coupled light-matter systems}},\ }\href {https://doi.org/10.1103/PhysRevLett.134.073002} {\bibfield  {journal} {\bibinfo  {journal} {Phys. Rev. Lett.}\ }\textbf {\bibinfo {volume} {134}},\ \bibinfo {pages} {73002} (\bibinfo {year} {2025})}\BibitemShut {NoStop}%
\bibitem [{\citenamefont {Fiscelli}\ \emph {et~al.}(2020)\citenamefont {Fiscelli}, \citenamefont {Rizzuto},\ and\ \citenamefont {Passante}}]{Fiscelli2020}%
  \BibitemOpen
  \bibfield  {author} {\bibinfo {author} {\bibfnamefont {G.}~\bibnamefont {Fiscelli}}, \bibinfo {author} {\bibfnamefont {L.}~\bibnamefont {Rizzuto}},\ and\ \bibinfo {author} {\bibfnamefont {R.}~\bibnamefont {Passante}},\ }\bibfield  {title} {\bibinfo {title} {Dispersion interaction between two hydrogen atoms in a static electric field},\ }\href {https://doi.org/10.1103/PhysRevLett.124.013604} {\bibfield  {journal} {\bibinfo  {journal} {Phys. Rev. Lett.}\ }\textbf {\bibinfo {volume} {124}},\ \bibinfo {pages} {013604} (\bibinfo {year} {2020})}\BibitemShut {NoStop}%
\bibitem [{\citenamefont {Karimpour}\ \emph {et~al.}(2022)\citenamefont {Karimpour}, \citenamefont {Fedorov},\ and\ \citenamefont {Tkatchenko}}]{Karimpour2022}%
  \BibitemOpen
  \bibfield  {author} {\bibinfo {author} {\bibfnamefont {M.~R.}\ \bibnamefont {Karimpour}}, \bibinfo {author} {\bibfnamefont {D.~V.}\ \bibnamefont {Fedorov}},\ and\ \bibinfo {author} {\bibfnamefont {A.}~\bibnamefont {Tkatchenko}},\ }\bibfield  {title} {\bibinfo {title} {{Quantum framework for describing retarded and nonretarded molecular interactions in external electric fields}},\ }\href {https://doi.org/10.1103/PhysRevResearch.4.013011} {\bibfield  {journal} {\bibinfo  {journal} {Phys. Rev. Res.}\ }\textbf {\bibinfo {volume} {4}},\ \bibinfo {pages} {013011} (\bibinfo {year} {2022})}\BibitemShut {NoStop}%
\bibitem [{\citenamefont {Saville}\ \emph {et~al.}(2006)\citenamefont {Saville}, \citenamefont {Chun}, \citenamefont {Li}, \citenamefont {Schniepp}, \citenamefont {Car},\ and\ \citenamefont {Aksay}}]{Saville2006}%
  \BibitemOpen
  \bibfield  {author} {\bibinfo {author} {\bibfnamefont {D.~A.}\ \bibnamefont {Saville}}, \bibinfo {author} {\bibfnamefont {J.}~\bibnamefont {Chun}}, \bibinfo {author} {\bibfnamefont {J.~L.}\ \bibnamefont {Li}}, \bibinfo {author} {\bibfnamefont {H.~C.}\ \bibnamefont {Schniepp}}, \bibinfo {author} {\bibfnamefont {R.}~\bibnamefont {Car}},\ and\ \bibinfo {author} {\bibfnamefont {I.~A.}\ \bibnamefont {Aksay}},\ }\bibfield  {title} {\bibinfo {title} {{Orientational order of molecular assemblies on inorganic crystals}},\ }\href {https://doi.org/10.1103/PhysRevLett.96.018301} {\bibfield  {journal} {\bibinfo  {journal} {Phys. Rev. Lett.}\ }\textbf {\bibinfo {volume} {96}},\ \bibinfo {pages} {018301} (\bibinfo {year} {2006})}\BibitemShut {NoStop}%
\bibitem [{\citenamefont {Emig}\ \emph {et~al.}(2009)\citenamefont {Emig}, \citenamefont {Graham}, \citenamefont {Jaffe},\ and\ \citenamefont {Kardar}}]{Emig2009}%
  \BibitemOpen
  \bibfield  {author} {\bibinfo {author} {\bibfnamefont {T.}~\bibnamefont {Emig}}, \bibinfo {author} {\bibfnamefont {N.}~\bibnamefont {Graham}}, \bibinfo {author} {\bibfnamefont {R.~L.}\ \bibnamefont {Jaffe}},\ and\ \bibinfo {author} {\bibfnamefont {M.}~\bibnamefont {Kardar}},\ }\bibfield  {title} {\bibinfo {title} {{Orientation dependence of Casimir forces}},\ }\href {https://doi.org/10.1103/PhysRevA.79.054901} {\bibfield  {journal} {\bibinfo  {journal} {Phys. Rev. A}\ }\textbf {\bibinfo {volume} {79}},\ \bibinfo {pages} {054901} (\bibinfo {year} {2009})}\BibitemShut {NoStop}%
\bibitem [{\citenamefont {Zhang}\ \emph {et~al.}(2017)\citenamefont {Zhang}, \citenamefont {He}, \citenamefont {Sushko}, \citenamefont {Liu}, \citenamefont {Luo}, \citenamefont {{De Yoreo}}, \citenamefont {Mao}, \citenamefont {Wang},\ and\ \citenamefont {Rosso}}]{Zhang2017}%
  \BibitemOpen
  \bibfield  {author} {\bibinfo {author} {\bibfnamefont {X.}~\bibnamefont {Zhang}}, \bibinfo {author} {\bibfnamefont {Y.}~\bibnamefont {He}}, \bibinfo {author} {\bibfnamefont {M.~L.}\ \bibnamefont {Sushko}}, \bibinfo {author} {\bibfnamefont {J.}~\bibnamefont {Liu}}, \bibinfo {author} {\bibfnamefont {L.}~\bibnamefont {Luo}}, \bibinfo {author} {\bibfnamefont {J.~J.}\ \bibnamefont {{De Yoreo}}}, \bibinfo {author} {\bibfnamefont {S.~X.}\ \bibnamefont {Mao}}, \bibinfo {author} {\bibfnamefont {C.}~\bibnamefont {Wang}},\ and\ \bibinfo {author} {\bibfnamefont {K.~M.}\ \bibnamefont {Rosso}},\ }\bibfield  {title} {\bibinfo {title} {{Direction-specific van der Waals attraction between rutile TiO2 nanocrystals}},\ }\href {https://doi.org/10.1126/science.aah6902} {\bibfield  {journal} {\bibinfo  {journal} {Science}\ }\textbf {\bibinfo {volume} {356}},\ \bibinfo {pages} {434} (\bibinfo {year} {2017})}\BibitemShut {NoStop}%
\bibitem [{\citenamefont {Lu}(2018)}]{Lu2018}%
  \BibitemOpen
  \bibfield  {author} {\bibinfo {author} {\bibfnamefont {B.~S.}\ \bibnamefont {Lu}},\ }\bibfield  {title} {\bibinfo {title} {{Van der Waals torque and force between anisotropic topological insulator slabs}},\ }\href {https://doi.org/10.1103/PhysRevB.97.045427} {\bibfield  {journal} {\bibinfo  {journal} {Phys. Rev. B}\ }\textbf {\bibinfo {volume} {97}},\ \bibinfo {pages} {045427} (\bibinfo {year} {2018})}\BibitemShut {NoStop}%
\bibitem [{\citenamefont {Wang}\ and\ \citenamefont {Antezza}(2024)}]{Wang2024}%
  \BibitemOpen
  \bibfield  {author} {\bibinfo {author} {\bibfnamefont {J.~S.}\ \bibnamefont {Wang}}\ and\ \bibinfo {author} {\bibfnamefont {M.}~\bibnamefont {Antezza}},\ }\bibfield  {title} {\bibinfo {title} {{Photon mediated transport of energy, linear momentum, and angular momentum in fullerene and graphene systems beyond local equilibrium}},\ }\href {https://link.aps.org/doi/10.1103/PhysRevB.109.125105} {\bibfield  {journal} {\bibinfo  {journal} {Phys. Rev. B}\ }\textbf {\bibinfo {volume} {109}} (\bibinfo {year} {2024})}\BibitemShut {NoStop}%
\bibitem [{\citenamefont {Kou}\ \emph {et~al.}(2024)\citenamefont {Kou}, \citenamefont {Zhou}, \citenamefont {Jiang}, \citenamefont {Tkatchenko},\ and\ \citenamefont {Liu}}]{Kou2024}%
  \BibitemOpen
  \bibfield  {author} {\bibinfo {author} {\bibfnamefont {Z.}~\bibnamefont {Kou}}, \bibinfo {author} {\bibfnamefont {Y.}~\bibnamefont {Zhou}}, \bibinfo {author} {\bibfnamefont {Z.}~\bibnamefont {Jiang}}, \bibinfo {author} {\bibfnamefont {A.}~\bibnamefont {Tkatchenko}},\ and\ \bibinfo {author} {\bibfnamefont {X.}~\bibnamefont {Liu}},\ }\href@noop {} {\bibinfo {title} {{van der Waals torque in 2D materials induced by interaction between many-body charge density fluctuations}}} (\bibinfo {year} {2024}),\ \Eprint {https://arxiv.org/abs/arxiv:2412.11357} {arxiv:2412.11357} \BibitemShut {NoStop}%
\bibitem [{\citenamefont {Shi}\ \emph {et~al.}(2015)\citenamefont {Shi}, \citenamefont {Hong}, \citenamefont {Bechtel}, \citenamefont {Zeng}, \citenamefont {Martin}, \citenamefont {Watanabe}, \citenamefont {Taniguchi}, \citenamefont {Shen},\ and\ \citenamefont {Wang}}]{Shi2015}%
  \BibitemOpen
  \bibfield  {author} {\bibinfo {author} {\bibfnamefont {Z.}~\bibnamefont {Shi}}, \bibinfo {author} {\bibfnamefont {X.}~\bibnamefont {Hong}}, \bibinfo {author} {\bibfnamefont {H.~A.}\ \bibnamefont {Bechtel}}, \bibinfo {author} {\bibfnamefont {B.}~\bibnamefont {Zeng}}, \bibinfo {author} {\bibfnamefont {M.~C.}\ \bibnamefont {Martin}}, \bibinfo {author} {\bibfnamefont {K.}~\bibnamefont {Watanabe}}, \bibinfo {author} {\bibfnamefont {T.}~\bibnamefont {Taniguchi}}, \bibinfo {author} {\bibfnamefont {Y.~R.}\ \bibnamefont {Shen}},\ and\ \bibinfo {author} {\bibfnamefont {F.}~\bibnamefont {Wang}},\ }\bibfield  {title} {\bibinfo {title} {{Observation of a Luttinger-liquid plasmon in metallic single-walled carbon nanotubes}},\ }\href {https://doi.org/10.1038/nphoton.2015.123} {\bibfield  {journal} {\bibinfo  {journal} {Nat. Photonics}\ }\textbf {\bibinfo {volume} {9}},\ \bibinfo {pages} {515} (\bibinfo {year} {2015})}\BibitemShut {NoStop}%
\bibitem [{\citenamefont {Wang}\ \emph {et~al.}(2022)\citenamefont {Wang}, \citenamefont {Yu}, \citenamefont {Kwan}, \citenamefont {Jia}, \citenamefont {Lei}, \citenamefont {Klemenz}, \citenamefont {Cevallos}, \citenamefont {Singha}, \citenamefont {Devakul}, \citenamefont {Watanabe}, \citenamefont {Taniguchi}, \citenamefont {Sondhi}, \citenamefont {Cava}, \citenamefont {Schoop}, \citenamefont {Parameswaran},\ and\ \citenamefont {Wu}}]{Wang2022}%
  \BibitemOpen
  \bibfield  {author} {\bibinfo {author} {\bibfnamefont {P.}~\bibnamefont {Wang}}, \bibinfo {author} {\bibfnamefont {G.}~\bibnamefont {Yu}}, \bibinfo {author} {\bibfnamefont {Y.~H.}\ \bibnamefont {Kwan}}, \bibinfo {author} {\bibfnamefont {Y.}~\bibnamefont {Jia}}, \bibinfo {author} {\bibfnamefont {S.}~\bibnamefont {Lei}}, \bibinfo {author} {\bibfnamefont {S.}~\bibnamefont {Klemenz}}, \bibinfo {author} {\bibfnamefont {F.~A.}\ \bibnamefont {Cevallos}}, \bibinfo {author} {\bibfnamefont {R.}~\bibnamefont {Singha}}, \bibinfo {author} {\bibfnamefont {T.}~\bibnamefont {Devakul}}, \bibinfo {author} {\bibfnamefont {K.}~\bibnamefont {Watanabe}}, \bibinfo {author} {\bibfnamefont {T.}~\bibnamefont {Taniguchi}}, \bibinfo {author} {\bibfnamefont {S.~L.}\ \bibnamefont {Sondhi}}, \bibinfo {author} {\bibfnamefont {R.~J.}\ \bibnamefont {Cava}}, \bibinfo {author} {\bibfnamefont {L.~M.}\ \bibnamefont {Schoop}}, \bibinfo {author} {\bibfnamefont {S.~A.}\ \bibnamefont {Parameswaran}},\ and\ \bibinfo {author} {\bibfnamefont
  {S.}~\bibnamefont {Wu}},\ }\bibfield  {title} {\bibinfo {title} {{One-dimensional Luttinger liquids in a two-dimensional moir{\'{e}} lattice}},\ }\href {https://doi.org/10.1038/s41586-022-04514-6} {\bibfield  {journal} {\bibinfo  {journal} {Nature}\ }\textbf {\bibinfo {volume} {605}},\ \bibinfo {pages} {57} (\bibinfo {year} {2022})}\BibitemShut {NoStop}%
\bibitem [{\citenamefont {Dobson}\ \emph {et~al.}(2009)\citenamefont {Dobson}, \citenamefont {Gould},\ and\ \citenamefont {Klich}}]{Dobson2009}%
  \BibitemOpen
  \bibfield  {author} {\bibinfo {author} {\bibfnamefont {J.~F.}\ \bibnamefont {Dobson}}, \bibinfo {author} {\bibfnamefont {T.}~\bibnamefont {Gould}},\ and\ \bibinfo {author} {\bibfnamefont {I.}~\bibnamefont {Klich}},\ }\bibfield  {title} {\bibinfo {title} {{Dispersion interaction between crossed conducting wires}},\ }\href {https://doi.org/10.1103/PhysRevA.80.012506} {\bibfield  {journal} {\bibinfo  {journal} {Phys. Rev. A}\ }\textbf {\bibinfo {volume} {80}},\ \bibinfo {pages} {012506} (\bibinfo {year} {2009})}\BibitemShut {NoStop}%
\bibitem [{\citenamefont {Rodriguez-Lopez}\ and\ \citenamefont {Emig}(2012)}]{Rodriguez-Lopez2012}%
  \BibitemOpen
  \bibfield  {author} {\bibinfo {author} {\bibfnamefont {P.}~\bibnamefont {Rodriguez-Lopez}}\ and\ \bibinfo {author} {\bibfnamefont {T.}~\bibnamefont {Emig}},\ }\bibfield  {title} {\bibinfo {title} {{Casimir interaction between inclined metallic cylinders}},\ }\href {https://doi.org/10.1103/PhysRevA.85.032510} {\bibfield  {journal} {\bibinfo  {journal} {Phys. Rev. A}\ }\textbf {\bibinfo {volume} {85}},\ \bibinfo {pages} {032510} (\bibinfo {year} {2012})}\BibitemShut {NoStop}%
\bibitem [{\citenamefont {White}\ and\ \citenamefont {Dobson}(2008)}]{White2008}%
  \BibitemOpen
  \bibfield  {author} {\bibinfo {author} {\bibfnamefont {A.}~\bibnamefont {White}}\ and\ \bibinfo {author} {\bibfnamefont {J.~F.}\ \bibnamefont {Dobson}},\ }\bibfield  {title} {\bibinfo {title} {{Enhanced dispersion interaction between quasi-one-dimensional conducting collinear structures}},\ }\href {https://doi.org/10.1103/PhysRevB.77.075436} {\bibfield  {journal} {\bibinfo  {journal} {Phys. Rev. B}\ }\textbf {\bibinfo {volume} {77}},\ \bibinfo {pages} {075436} (\bibinfo {year} {2008})}\BibitemShut {NoStop}%
\bibitem [{\citenamefont {Misquitta}\ \emph {et~al.}(2010)\citenamefont {Misquitta}, \citenamefont {Spencer}, \citenamefont {Stone},\ and\ \citenamefont {Alavi}}]{Misquitta2010}%
  \BibitemOpen
  \bibfield  {author} {\bibinfo {author} {\bibfnamefont {A.~J.}\ \bibnamefont {Misquitta}}, \bibinfo {author} {\bibfnamefont {J.}~\bibnamefont {Spencer}}, \bibinfo {author} {\bibfnamefont {A.~J.}\ \bibnamefont {Stone}},\ and\ \bibinfo {author} {\bibfnamefont {A.}~\bibnamefont {Alavi}},\ }\bibfield  {title} {\bibinfo {title} {{Dispersion interactions between semiconducting wires}},\ }\href {https://doi.org/10.1103/PhysRevB.82.075312} {\bibfield  {journal} {\bibinfo  {journal} {Phys. Rev. B}\ }\textbf {\bibinfo {volume} {82}},\ \bibinfo {pages} {075312} (\bibinfo {year} {2010})}\BibitemShut {NoStop}%
\bibitem [{\citenamefont {Misquitta}\ \emph {et~al.}(2014)\citenamefont {Misquitta}, \citenamefont {Maezono}, \citenamefont {Drummond}, \citenamefont {Stone},\ and\ \citenamefont {Needs}}]{Misquitta2014}%
  \BibitemOpen
  \bibfield  {author} {\bibinfo {author} {\bibfnamefont {A.~J.}\ \bibnamefont {Misquitta}}, \bibinfo {author} {\bibfnamefont {R.}~\bibnamefont {Maezono}}, \bibinfo {author} {\bibfnamefont {N.~D.}\ \bibnamefont {Drummond}}, \bibinfo {author} {\bibfnamefont {A.~J.}\ \bibnamefont {Stone}},\ and\ \bibinfo {author} {\bibfnamefont {R.~J.}\ \bibnamefont {Needs}},\ }\bibfield  {title} {\bibinfo {title} {{Anomalous nonadditive dispersion interactions in systems of three one-dimensional wires}},\ }\href {https://doi.org/10.1103/PhysRevB.89.045140} {\bibfield  {journal} {\bibinfo  {journal} {Phys. Rev. B}\ }\textbf {\bibinfo {volume} {89}},\ \bibinfo {pages} {045140} (\bibinfo {year} {2014})}\BibitemShut {NoStop}%
\bibitem [{\citenamefont {{De Vega}}\ \emph {et~al.}(2016)\citenamefont {{De Vega}}, \citenamefont {Cox},\ and\ \citenamefont {{De Abajo}}}]{DeVega2016}%
  \BibitemOpen
  \bibfield  {author} {\bibinfo {author} {\bibfnamefont {S.}~\bibnamefont {{De Vega}}}, \bibinfo {author} {\bibfnamefont {J.~D.}\ \bibnamefont {Cox}},\ and\ \bibinfo {author} {\bibfnamefont {F.~J.~G.}\ \bibnamefont {{De Abajo}}},\ }\bibfield  {title} {\bibinfo {title} {{Plasmons in doped finite carbon nanotubes and their interactions with fast electrons and quantum emitters}},\ }\href {https://doi.org/10.1103/PhysRevB.94.075447} {\bibfield  {journal} {\bibinfo  {journal} {Phys. Rev. B}\ }\textbf {\bibinfo {volume} {94}},\ \bibinfo {pages} {075447} (\bibinfo {year} {2016})}\BibitemShut {NoStop}%
\bibitem [{\citenamefont {Wang}\ \emph {et~al.}(2023)\citenamefont {Wang}, \citenamefont {Peng}, \citenamefont {Zhang}, \citenamefont {Zhang},\ and\ \citenamefont {Zhu}}]{Wang2023}%
  \BibitemOpen
  \bibfield  {author} {\bibinfo {author} {\bibfnamefont {J.-S.}\ \bibnamefont {Wang}}, \bibinfo {author} {\bibfnamefont {J.}~\bibnamefont {Peng}}, \bibinfo {author} {\bibfnamefont {Z.-Q.}\ \bibnamefont {Zhang}}, \bibinfo {author} {\bibfnamefont {Y.-M.}\ \bibnamefont {Zhang}},\ and\ \bibinfo {author} {\bibfnamefont {T.}~\bibnamefont {Zhu}},\ }\bibfield  {title} {\bibinfo {title} {{Transport in electron-photon systems}},\ }\href {https://doi.org/10.1007/s11467-023-1260-z} {\bibfield  {journal} {\bibinfo  {journal} {Front. Phys.}\ }\textbf {\bibinfo {volume} {18}},\ \bibinfo {pages} {43602} (\bibinfo {year} {2023})}\BibitemShut {NoStop}%
\bibitem [{\citenamefont {Chudnovskiy}\ \emph {et~al.}(2023)\citenamefont {Chudnovskiy}, \citenamefont {Levchenko},\ and\ \citenamefont {Kamenev}}]{Chudnovskiy2023}%
  \BibitemOpen
  \bibfield  {author} {\bibinfo {author} {\bibfnamefont {A.~L.}\ \bibnamefont {Chudnovskiy}}, \bibinfo {author} {\bibfnamefont {A.}~\bibnamefont {Levchenko}},\ and\ \bibinfo {author} {\bibfnamefont {A.}~\bibnamefont {Kamenev}},\ }\bibfield  {title} {\bibinfo {title} {{Coulomb Drag and Heat Transfer in Strange Metals}},\ }\href {https://doi.org/10.1103/PhysRevLett.131.096501} {\bibfield  {journal} {\bibinfo  {journal} {Phys. Rev. Lett.}\ }\textbf {\bibinfo {volume} {131}},\ \bibinfo {pages} {096501} (\bibinfo {year} {2023})}\BibitemShut {NoStop}%
\bibitem [{\citenamefont {Pan}\ \emph {et~al.}(2024)\citenamefont {Pan}, \citenamefont {Ren}, \citenamefont {Tang},\ and\ \citenamefont {Wang}}]{Pan2024}%
  \BibitemOpen
  \bibfield  {author} {\bibinfo {author} {\bibfnamefont {H.}~\bibnamefont {Pan}}, \bibinfo {author} {\bibfnamefont {Y.}~\bibnamefont {Ren}}, \bibinfo {author} {\bibfnamefont {G.}~\bibnamefont {Tang}},\ and\ \bibinfo {author} {\bibfnamefont {J.-S.}\ \bibnamefont {Wang}},\ }\href@noop {} {\bibinfo {title} {{Asymmetry-induced radiative heat transfer in Floquet systems}}} (\bibinfo {year} {2024}),\ \Eprint {https://arxiv.org/abs/2410.10176} {arXiv:2410.10176} \BibitemShut {NoStop}%
\bibitem [{\citenamefont {Dobson}\ and\ \citenamefont {Gould}(2012)}]{Dobson2012}%
  \BibitemOpen
  \bibfield  {author} {\bibinfo {author} {\bibfnamefont {J.~F.}\ \bibnamefont {Dobson}}\ and\ \bibinfo {author} {\bibfnamefont {T.}~\bibnamefont {Gould}},\ }\bibfield  {title} {\bibinfo {title} {{Calculation of dispersion energies}},\ }\href {https://doi.org/10.1088/0953-8984/24/7/073201} {\bibfield  {journal} {\bibinfo  {journal} {J. Phys. Condens. Matter}\ }\textbf {\bibinfo {volume} {24}},\ \bibinfo {pages} {073201} (\bibinfo {year} {2012})}\BibitemShut {NoStop}%
\bibitem [{\citenamefont {Toulouse}\ \emph {et~al.}(2009)\citenamefont {Toulouse}, \citenamefont {Gerber}, \citenamefont {Jansen}, \citenamefont {Savin},\ and\ \citenamefont {{\'{A}}ngy{\'{a}}n}}]{Toulouse2009}%
  \BibitemOpen
  \bibfield  {author} {\bibinfo {author} {\bibfnamefont {J.}~\bibnamefont {Toulouse}}, \bibinfo {author} {\bibfnamefont {I.~C.}\ \bibnamefont {Gerber}}, \bibinfo {author} {\bibfnamefont {G.}~\bibnamefont {Jansen}}, \bibinfo {author} {\bibfnamefont {A.}~\bibnamefont {Savin}},\ and\ \bibinfo {author} {\bibfnamefont {J.~G.}\ \bibnamefont {{\'{A}}ngy{\'{a}}n}},\ }\bibfield  {title} {\bibinfo {title} {{Adiabatic-connection fluctuation-dissipation density-functional theory based on range separation}},\ }\href {https://doi.org/10.1103/PhysRevLett.102.096404} {\bibfield  {journal} {\bibinfo  {journal} {Phys. Rev. Lett.}\ }\textbf {\bibinfo {volume} {102}},\ \bibinfo {pages} {096404} (\bibinfo {year} {2009})}\BibitemShut {NoStop}%
\bibitem [{\citenamefont {Tkatchenko}\ \emph {et~al.}(2013)\citenamefont {Tkatchenko}, \citenamefont {Ambrosetti},\ and\ \citenamefont {DiStasio}}]{Tkatchenko2013}%
  \BibitemOpen
  \bibfield  {author} {\bibinfo {author} {\bibfnamefont {A.}~\bibnamefont {Tkatchenko}}, \bibinfo {author} {\bibfnamefont {A.}~\bibnamefont {Ambrosetti}},\ and\ \bibinfo {author} {\bibfnamefont {R.~A.}\ \bibnamefont {DiStasio}},\ }\bibfield  {title} {\bibinfo {title} {{Interatomic methods for the dispersion energy derived from the adiabatic connection fluctuation-dissipation theorem}},\ }\href {https://doi.org/10.1063/1.4789814} {\bibfield  {journal} {\bibinfo  {journal} {J. Chem. Phys.}\ }\textbf {\bibinfo {volume} {138}} (\bibinfo {year} {2013})}\BibitemShut {NoStop}%
\bibitem [{\citenamefont {Drosdoff}\ and\ \citenamefont {Woods}(2014)}]{Drosdoff2014}%
  \BibitemOpen
  \bibfield  {author} {\bibinfo {author} {\bibfnamefont {D.}~\bibnamefont {Drosdoff}}\ and\ \bibinfo {author} {\bibfnamefont {L.~M.}\ \bibnamefont {Woods}},\ }\bibfield  {title} {\bibinfo {title} {{Quantum and thermal dispersion forces: Application to graphene nanoribbons}},\ }\href {https://doi.org/10.1103/physrevlett.112.025501} {\bibfield  {journal} {\bibinfo  {journal} {Phys. Rev. Lett.}\ }\textbf {\bibinfo {volume} {112}},\ \bibinfo {pages} {025501} (\bibinfo {year} {2014})}\BibitemShut {NoStop}%
\bibitem [{\citenamefont {Ramberger}\ \emph {et~al.}(2017)\citenamefont {Ramberger}, \citenamefont {Sch{\"{a}}fer},\ and\ \citenamefont {Kresse}}]{Ramberger2017}%
  \BibitemOpen
  \bibfield  {author} {\bibinfo {author} {\bibfnamefont {B.}~\bibnamefont {Ramberger}}, \bibinfo {author} {\bibfnamefont {T.}~\bibnamefont {Sch{\"{a}}fer}},\ and\ \bibinfo {author} {\bibfnamefont {G.}~\bibnamefont {Kresse}},\ }\bibfield  {title} {\bibinfo {title} {{Analytic interatomic forces in the random phase approximation}},\ }\href {https://doi.org/10.1103/PhysRevLett.118.106403} {\bibfield  {journal} {\bibinfo  {journal} {Phys. Rev. Lett.}\ }\textbf {\bibinfo {volume} {118}},\ \bibinfo {pages} {106403} (\bibinfo {year} {2017})}\BibitemShut {NoStop}%
\bibitem [{\citenamefont {Jackson}(1999)}]{jackson1999classical}%
  \BibitemOpen
  \bibfield  {author} {\bibinfo {author} {\bibfnamefont {J.~D.}\ \bibnamefont {Jackson}},\ }\href@noop {} {\emph {\bibinfo {title} {{Classical Electrodynamics}}}},\ \bibinfo {edition} {3rd}\ ed.\ (\bibinfo  {publisher} {John Wiley \& Sons},\ \bibinfo {year} {1999})\BibitemShut {NoStop}%
\bibitem [{\citenamefont {Settnes}\ \emph {et~al.}(2017)\citenamefont {Settnes}, \citenamefont {Saavedra}, \citenamefont {Thygesen}, \citenamefont {Jauho}, \citenamefont {{Garc{\'{i}}a De Abajo}},\ and\ \citenamefont {Mortensen}}]{Settnes2017}%
  \BibitemOpen
  \bibfield  {author} {\bibinfo {author} {\bibfnamefont {M.}~\bibnamefont {Settnes}}, \bibinfo {author} {\bibfnamefont {J.~R.}\ \bibnamefont {Saavedra}}, \bibinfo {author} {\bibfnamefont {K.~S.}\ \bibnamefont {Thygesen}}, \bibinfo {author} {\bibfnamefont {A.~P.}\ \bibnamefont {Jauho}}, \bibinfo {author} {\bibfnamefont {F.~J.}\ \bibnamefont {{Garc{\'{i}}a De Abajo}}},\ and\ \bibinfo {author} {\bibfnamefont {N.~A.}\ \bibnamefont {Mortensen}},\ }\bibfield  {title} {\bibinfo {title} {{Strong plasmon-phonon splitting and hybridization in 2D materials revealed through a self-energy approach}},\ }\href {https://doi.org/10.1021/acsphotonics.7b00928} {\bibfield  {journal} {\bibinfo  {journal} {ACS Photonics}\ }\textbf {\bibinfo {volume} {4}},\ \bibinfo {pages} {2908} (\bibinfo {year} {2017})}\BibitemShut {NoStop}%
\bibitem [{\citenamefont {Li}\ and\ \citenamefont {{Das Sarma}}(1989)}]{Li1989}%
  \BibitemOpen
  \bibfield  {author} {\bibinfo {author} {\bibfnamefont {Q.}~\bibnamefont {Li}}\ and\ \bibinfo {author} {\bibfnamefont {S.}~\bibnamefont {{Das Sarma}}},\ }\bibfield  {title} {\bibinfo {title} {{Collective excitation spectra of one-dimensional electron systems}},\ }\href {https://doi.org/10.1103/PhysRevB.40.5860} {\bibfield  {journal} {\bibinfo  {journal} {Phys. Rev. B}\ }\textbf {\bibinfo {volume} {40}},\ \bibinfo {pages} {5860} (\bibinfo {year} {1989})}\BibitemShut {NoStop}%
\bibitem [{SM()}]{SM}%
  \BibitemOpen
  \href@noop {} {}\bibinfo {note} {{See Supplemental Material for details of Green's function method, scaling behaviors in carbon nanotubes, numerical parameter settings, as well as analyses based on Luttinger liquid theory and dipole model, which includes Refs.~\cite{Ozaki2003,Bucko2016,Hermann2023}.}}\BibitemShut {Stop}%
\bibitem [{\citenamefont {Grimme}\ \emph {et~al.}(2010)\citenamefont {Grimme}, \citenamefont {Antony}, \citenamefont {Ehrlich},\ and\ \citenamefont {Krieg}}]{Grimme2010}%
  \BibitemOpen
  \bibfield  {author} {\bibinfo {author} {\bibfnamefont {S.}~\bibnamefont {Grimme}}, \bibinfo {author} {\bibfnamefont {J.}~\bibnamefont {Antony}}, \bibinfo {author} {\bibfnamefont {S.}~\bibnamefont {Ehrlich}},\ and\ \bibinfo {author} {\bibfnamefont {H.}~\bibnamefont {Krieg}},\ }\bibfield  {title} {\bibinfo {title} {A consistent and accurate ab initio parametrization of density functional dispersion correction (dft-d) for the 94 elements h-pu},\ }\href {https://doi.org/10.1063/1.3382344} {\bibfield  {journal} {\bibinfo  {journal} {J. Chem. Phys.}\ }\textbf {\bibinfo {volume} {132}},\ \bibinfo {pages} {154104} (\bibinfo {year} {2010})}\BibitemShut {NoStop}%
\bibitem [{\citenamefont {Nagao}\ \emph {et~al.}(2006)\citenamefont {Nagao}, \citenamefont {Yaginuma}, \citenamefont {Inaoka},\ and\ \citenamefont {Sakurai}}]{Nagao2006}%
  \BibitemOpen
  \bibfield  {author} {\bibinfo {author} {\bibfnamefont {T.}~\bibnamefont {Nagao}}, \bibinfo {author} {\bibfnamefont {S.}~\bibnamefont {Yaginuma}}, \bibinfo {author} {\bibfnamefont {T.}~\bibnamefont {Inaoka}},\ and\ \bibinfo {author} {\bibfnamefont {T.}~\bibnamefont {Sakurai}},\ }\bibfield  {title} {\bibinfo {title} {{One-dimensional plasmon in an atomic-scale metal wire}},\ }\href {https://doi.org/10.1103/PhysRevLett.97.116802} {\bibfield  {journal} {\bibinfo  {journal} {Phys. Rev. Lett.}\ }\textbf {\bibinfo {volume} {97}},\ \bibinfo {pages} {116802} (\bibinfo {year} {2006})}\BibitemShut {NoStop}%
\bibitem [{\citenamefont {Dobson}\ and\ \citenamefont {Ambrosetti}(2023)}]{Dobson2023}%
  \BibitemOpen
  \bibfield  {author} {\bibinfo {author} {\bibfnamefont {J.~F.}\ \bibnamefont {Dobson}}\ and\ \bibinfo {author} {\bibfnamefont {A.}~\bibnamefont {Ambrosetti}},\ }\bibfield  {title} {\bibinfo {title} {{MBD + C: How to incorporate metallic character into atom-based dispersion energy schemes}},\ }\href {https://doi.org/10.1021/acs.jctc.3c00353} {\bibfield  {journal} {\bibinfo  {journal} {J. Chem. Theory Comput.}\ }\textbf {\bibinfo {volume} {19}},\ \bibinfo {pages} {6434} (\bibinfo {year} {2023})}\BibitemShut {NoStop}%
\bibitem [{\citenamefont {Yu}\ \emph {et~al.}(2023)\citenamefont {Yu}, \citenamefont {Wang}, \citenamefont {Uzan-Narovlansky}, \citenamefont {Jia}, \citenamefont {Onyszczak}, \citenamefont {Singha}, \citenamefont {Gui}, \citenamefont {Song}, \citenamefont {Tang}, \citenamefont {Watanabe}, \citenamefont {Taniguchi}, \citenamefont {Cava}, \citenamefont {Schoop},\ and\ \citenamefont {Wu}}]{Yu2023}%
  \BibitemOpen
  \bibfield  {author} {\bibinfo {author} {\bibfnamefont {G.}~\bibnamefont {Yu}}, \bibinfo {author} {\bibfnamefont {P.}~\bibnamefont {Wang}}, \bibinfo {author} {\bibfnamefont {A.~J.}\ \bibnamefont {Uzan-Narovlansky}}, \bibinfo {author} {\bibfnamefont {Y.}~\bibnamefont {Jia}}, \bibinfo {author} {\bibfnamefont {M.}~\bibnamefont {Onyszczak}}, \bibinfo {author} {\bibfnamefont {R.}~\bibnamefont {Singha}}, \bibinfo {author} {\bibfnamefont {X.}~\bibnamefont {Gui}}, \bibinfo {author} {\bibfnamefont {T.}~\bibnamefont {Song}}, \bibinfo {author} {\bibfnamefont {Y.}~\bibnamefont {Tang}}, \bibinfo {author} {\bibfnamefont {K.}~\bibnamefont {Watanabe}}, \bibinfo {author} {\bibfnamefont {T.}~\bibnamefont {Taniguchi}}, \bibinfo {author} {\bibfnamefont {R.~J.}\ \bibnamefont {Cava}}, \bibinfo {author} {\bibfnamefont {L.~M.}\ \bibnamefont {Schoop}},\ and\ \bibinfo {author} {\bibfnamefont {S.}~\bibnamefont {Wu}},\ }\bibfield  {title} {\bibinfo {title} {{Evidence for two dimensional anisotropic Luttinger liquids at millikelvin
  temperatures}},\ }\href {https://doi.org/10.1038/s41467-023-42821-2} {\bibfield  {journal} {\bibinfo  {journal} {Nat. Commun.}\ }\textbf {\bibinfo {volume} {14}},\ \bibinfo {pages} {7025} (\bibinfo {year} {2023})}\BibitemShut {NoStop}%
\bibitem [{\citenamefont {Casimir}\ and\ \citenamefont {Polder}(1948)}]{Casimir1948}%
  \BibitemOpen
  \bibfield  {author} {\bibinfo {author} {\bibfnamefont {H.~B.~G.}\ \bibnamefont {Casimir}}\ and\ \bibinfo {author} {\bibfnamefont {D.}~\bibnamefont {Polder}},\ }\bibfield  {title} {\bibinfo {title} {{The influence of retardation on the London-van der Waals forces}},\ }\href {https://doi.org/10.1103/PhysRev.73.360} {\bibfield  {journal} {\bibinfo  {journal} {Phys. Rev.}\ }\textbf {\bibinfo {volume} {73}},\ \bibinfo {pages} {360} (\bibinfo {year} {1948})}\BibitemShut {NoStop}%
\bibitem [{\citenamefont {Zeybek}\ \emph {et~al.}(2023)\citenamefont {Zeybek}, \citenamefont {Mukherjee},\ and\ \citenamefont {Schmelcher}}]{Zeybek2023}%
  \BibitemOpen
  \bibfield  {author} {\bibinfo {author} {\bibfnamefont {Z.}~\bibnamefont {Zeybek}}, \bibinfo {author} {\bibfnamefont {R.}~\bibnamefont {Mukherjee}},\ and\ \bibinfo {author} {\bibfnamefont {P.}~\bibnamefont {Schmelcher}},\ }\bibfield  {title} {\bibinfo {title} {{Quantum Phases from Competing Van der Waals and Dipole-Dipole Interactions of Rydberg Atoms}},\ }\href {https://doi.org/10.1103/PhysRevLett.131.203003} {\bibfield  {journal} {\bibinfo  {journal} {Phys. Rev. Lett.}\ }\textbf {\bibinfo {volume} {131}},\ \bibinfo {pages} {203003} (\bibinfo {year} {2023})}\BibitemShut {NoStop}%
\bibitem [{\citenamefont {Yang}\ \emph {et~al.}(2017)\citenamefont {Yang}, \citenamefont {Chen}, \citenamefont {Zheng}, \citenamefont {Sun}, \citenamefont {Dai}, \citenamefont {Guan}, \citenamefont {Yuan},\ and\ \citenamefont {Pan}}]{Yang2017}%
  \BibitemOpen
  \bibfield  {author} {\bibinfo {author} {\bibfnamefont {B.}~\bibnamefont {Yang}}, \bibinfo {author} {\bibfnamefont {Y.-Y.}\ \bibnamefont {Chen}}, \bibinfo {author} {\bibfnamefont {Y.-G.}\ \bibnamefont {Zheng}}, \bibinfo {author} {\bibfnamefont {H.}~\bibnamefont {Sun}}, \bibinfo {author} {\bibfnamefont {H.-N.}\ \bibnamefont {Dai}}, \bibinfo {author} {\bibfnamefont {X.-W.}\ \bibnamefont {Guan}}, \bibinfo {author} {\bibfnamefont {Z.-S.}\ \bibnamefont {Yuan}},\ and\ \bibinfo {author} {\bibfnamefont {J.-W.}\ \bibnamefont {Pan}},\ }\bibfield  {title} {\bibinfo {title} {{Quantum criticality and the Tomonaga-Luttinger liquid in one-dimensional Bose gases}},\ }\href {https://doi.org/10.1103/PhysRevLett.119.165701} {\bibfield  {journal} {\bibinfo  {journal} {Phys. Rev. Lett.}\ }\textbf {\bibinfo {volume} {119}},\ \bibinfo {pages} {165701} (\bibinfo {year} {2017})}\BibitemShut {NoStop}%
\bibitem [{\citenamefont {Intravaia}\ \emph {et~al.}(2013)\citenamefont {Intravaia}, \citenamefont {Koev}, \citenamefont {Jung}, \citenamefont {Talin}, \citenamefont {Davids}, \citenamefont {Decca}, \citenamefont {Aksyuk}, \citenamefont {Dalvit},\ and\ \citenamefont {L{\'{o}}pez}}]{Intravaia2013}%
  \BibitemOpen
  \bibfield  {author} {\bibinfo {author} {\bibfnamefont {F.}~\bibnamefont {Intravaia}}, \bibinfo {author} {\bibfnamefont {S.}~\bibnamefont {Koev}}, \bibinfo {author} {\bibfnamefont {I.~W.}\ \bibnamefont {Jung}}, \bibinfo {author} {\bibfnamefont {A.~A.}\ \bibnamefont {Talin}}, \bibinfo {author} {\bibfnamefont {P.~S.}\ \bibnamefont {Davids}}, \bibinfo {author} {\bibfnamefont {R.~S.}\ \bibnamefont {Decca}}, \bibinfo {author} {\bibfnamefont {V.~A.}\ \bibnamefont {Aksyuk}}, \bibinfo {author} {\bibfnamefont {D.~A.~R.}\ \bibnamefont {Dalvit}},\ and\ \bibinfo {author} {\bibfnamefont {D.}~\bibnamefont {L{\'{o}}pez}},\ }\bibfield  {title} {\bibinfo {title} {{Strong Casimir force reduction through metallic surface nanostructuring}},\ }\href {https://doi.org/10.1038/ncomms3515} {\bibfield  {journal} {\bibinfo  {journal} {Nat. Commun.}\ }\textbf {\bibinfo {volume} {4}},\ \bibinfo {pages} {2515} (\bibinfo {year} {2013})}\BibitemShut {NoStop}%
\bibitem [{\citenamefont {Li}\ \emph {et~al.}(2019)\citenamefont {Li}, \citenamefont {Yin}, \citenamefont {Liu}, \citenamefont {Wu}, \citenamefont {Li}, \citenamefont {Li},\ and\ \citenamefont {Guo}}]{Li2019}%
  \BibitemOpen
  \bibfield  {author} {\bibinfo {author} {\bibfnamefont {B.}~\bibnamefont {Li}}, \bibinfo {author} {\bibfnamefont {J.}~\bibnamefont {Yin}}, \bibinfo {author} {\bibfnamefont {X.}~\bibnamefont {Liu}}, \bibinfo {author} {\bibfnamefont {H.}~\bibnamefont {Wu}}, \bibinfo {author} {\bibfnamefont {J.}~\bibnamefont {Li}}, \bibinfo {author} {\bibfnamefont {X.}~\bibnamefont {Li}},\ and\ \bibinfo {author} {\bibfnamefont {W.}~\bibnamefont {Guo}},\ }\bibfield  {title} {\bibinfo {title} {{Probing van der Waals interactions at two-dimensional heterointerfaces}},\ }\href {https://doi.org/10.1038/s41565-019-0405-2} {\bibfield  {journal} {\bibinfo  {journal} {Nat. Nanotechnol.}\ }\textbf {\bibinfo {volume} {14}},\ \bibinfo {pages} {567} (\bibinfo {year} {2019})}\BibitemShut {NoStop}%
\bibitem [{\citenamefont {Wang}\ \emph {et~al.}(2008)\citenamefont {Wang}, \citenamefont {Hu}, \citenamefont {Lieber},\ and\ \citenamefont {Sun}}]{Wang2008}%
  \BibitemOpen
  \bibfield  {author} {\bibinfo {author} {\bibfnamefont {C.}~\bibnamefont {Wang}}, \bibinfo {author} {\bibfnamefont {Y.}~\bibnamefont {Hu}}, \bibinfo {author} {\bibfnamefont {C.~M.}\ \bibnamefont {Lieber}},\ and\ \bibinfo {author} {\bibfnamefont {S.}~\bibnamefont {Sun}},\ }\bibfield  {title} {\bibinfo {title} {{Ultrathin Au Nanowires and Their Transport Properties}},\ }\href {https://doi.org/10.1021/ja803408f} {\bibfield  {journal} {\bibinfo  {journal} {J. Am. Chem. Soc.}\ }\textbf {\bibinfo {volume} {130}},\ \bibinfo {pages} {8902} (\bibinfo {year} {2008})}\BibitemShut {NoStop}%
\bibitem [{\citenamefont {Gao}\ \emph {et~al.}(2024)\citenamefont {Gao}, \citenamefont {Zheng}, \citenamefont {Sun}, \citenamefont {Kang}, \citenamefont {Zhou},\ and\ \citenamefont {Xu}}]{Gao2024}%
  \BibitemOpen
  \bibfield  {author} {\bibinfo {author} {\bibfnamefont {W.}~\bibnamefont {Gao}}, \bibinfo {author} {\bibfnamefont {W.}~\bibnamefont {Zheng}}, \bibinfo {author} {\bibfnamefont {L.}~\bibnamefont {Sun}}, \bibinfo {author} {\bibfnamefont {F.}~\bibnamefont {Kang}}, \bibinfo {author} {\bibfnamefont {Z.}~\bibnamefont {Zhou}},\ and\ \bibinfo {author} {\bibfnamefont {W.}~\bibnamefont {Xu}},\ }\bibfield  {title} {\bibinfo {title} {{On-surface synthesis and characterization of polyynic carbon chains}},\ }\href {https://academic.oup.com/nsr/article/doi/10.1093/nsr/nwae031/7585363} {\bibfield  {journal} {\bibinfo  {journal} {Natl. Sci. Rev.}\ }\textbf {\bibinfo {volume} {11}} (\bibinfo {year} {2024})}\BibitemShut {NoStop}%
\bibitem [{\citenamefont {Wang}\ \emph {et~al.}(2019)\citenamefont {Wang}, \citenamefont {Chen}, \citenamefont {Zhu}, \citenamefont {Wang}, \citenamefont {Dong}, \citenamefont {Sun}, \citenamefont {Zhang}, \citenamefont {Cao}, \citenamefont {Li}, \citenamefont {Huang}, \citenamefont {Zhang}, \citenamefont {Liu}, \citenamefont {Sun}, \citenamefont {Ye}, \citenamefont {Song}, \citenamefont {Wang}, \citenamefont {Han}, \citenamefont {Yang}, \citenamefont {Guo}, \citenamefont {Qin}, \citenamefont {Xiao}, \citenamefont {Zhang}, \citenamefont {Chen}, \citenamefont {Han},\ and\ \citenamefont {Zhang}}]{Wang2019}%
  \BibitemOpen
  \bibfield  {author} {\bibinfo {author} {\bibfnamefont {H.}~\bibnamefont {Wang}}, \bibinfo {author} {\bibfnamefont {M.~L.}\ \bibnamefont {Chen}}, \bibinfo {author} {\bibfnamefont {M.}~\bibnamefont {Zhu}}, \bibinfo {author} {\bibfnamefont {Y.}~\bibnamefont {Wang}}, \bibinfo {author} {\bibfnamefont {B.}~\bibnamefont {Dong}}, \bibinfo {author} {\bibfnamefont {X.}~\bibnamefont {Sun}}, \bibinfo {author} {\bibfnamefont {X.}~\bibnamefont {Zhang}}, \bibinfo {author} {\bibfnamefont {S.}~\bibnamefont {Cao}}, \bibinfo {author} {\bibfnamefont {X.}~\bibnamefont {Li}}, \bibinfo {author} {\bibfnamefont {J.}~\bibnamefont {Huang}}, \bibinfo {author} {\bibfnamefont {L.}~\bibnamefont {Zhang}}, \bibinfo {author} {\bibfnamefont {W.}~\bibnamefont {Liu}}, \bibinfo {author} {\bibfnamefont {D.}~\bibnamefont {Sun}}, \bibinfo {author} {\bibfnamefont {Y.}~\bibnamefont {Ye}}, \bibinfo {author} {\bibfnamefont {K.}~\bibnamefont {Song}}, \bibinfo {author} {\bibfnamefont {J.}~\bibnamefont {Wang}}, \bibinfo {author} {\bibfnamefont
  {Y.}~\bibnamefont {Han}}, \bibinfo {author} {\bibfnamefont {T.}~\bibnamefont {Yang}}, \bibinfo {author} {\bibfnamefont {H.}~\bibnamefont {Guo}}, \bibinfo {author} {\bibfnamefont {C.}~\bibnamefont {Qin}}, \bibinfo {author} {\bibfnamefont {L.}~\bibnamefont {Xiao}}, \bibinfo {author} {\bibfnamefont {J.}~\bibnamefont {Zhang}}, \bibinfo {author} {\bibfnamefont {J.}~\bibnamefont {Chen}}, \bibinfo {author} {\bibfnamefont {Z.}~\bibnamefont {Han}},\ and\ \bibinfo {author} {\bibfnamefont {Z.}~\bibnamefont {Zhang}},\ }\bibfield  {title} {\bibinfo {title} {{Gate tunable giant anisotropic resistance in ultra-thin GaTe}},\ }\href {https://doi.org/10.1038/s41467-019-10256-3} {\bibfield  {journal} {\bibinfo  {journal} {Nat. Commun.}\ }\textbf {\bibinfo {volume} {10}},\ \bibinfo {pages} {2302} (\bibinfo {year} {2019})}\BibitemShut {NoStop}%
\bibitem [{\citenamefont {Zhang}\ \emph {et~al.}(2025)\citenamefont {Zhang}, \citenamefont {Chen}, \citenamefont {Shen}, \citenamefont {Chen}, \citenamefont {Wang}, \citenamefont {Wang}, \citenamefont {Ma}, \citenamefont {Lyu}, \citenamefont {Zhou}, \citenamefont {Lou}, \citenamefont {Wu}, \citenamefont {Xie}, \citenamefont {Zhang}, \citenamefont {Wang}, \citenamefont {Xu}, \citenamefont {Li}, \citenamefont {Wang}, \citenamefont {Watanabe}, \citenamefont {Taniguchi}, \citenamefont {Qian}, \citenamefont {Jia}, \citenamefont {Liang}, \citenamefont {Wang}, \citenamefont {Yang}, \citenamefont {Zhang}, \citenamefont {Jin}, \citenamefont {Ouyang},\ and\ \citenamefont {Shi}}]{Zhang2025}%
  \BibitemOpen
  \bibfield  {author} {\bibinfo {author} {\bibfnamefont {Z.}~\bibnamefont {Zhang}}, \bibinfo {author} {\bibfnamefont {Y.}~\bibnamefont {Chen}}, \bibinfo {author} {\bibfnamefont {P.}~\bibnamefont {Shen}}, \bibinfo {author} {\bibfnamefont {J.}~\bibnamefont {Chen}}, \bibinfo {author} {\bibfnamefont {S.}~\bibnamefont {Wang}}, \bibinfo {author} {\bibfnamefont {B.}~\bibnamefont {Wang}}, \bibinfo {author} {\bibfnamefont {S.}~\bibnamefont {Ma}}, \bibinfo {author} {\bibfnamefont {B.}~\bibnamefont {Lyu}}, \bibinfo {author} {\bibfnamefont {X.}~\bibnamefont {Zhou}}, \bibinfo {author} {\bibfnamefont {S.}~\bibnamefont {Lou}}, \bibinfo {author} {\bibfnamefont {Z.}~\bibnamefont {Wu}}, \bibinfo {author} {\bibfnamefont {Y.}~\bibnamefont {Xie}}, \bibinfo {author} {\bibfnamefont {C.}~\bibnamefont {Zhang}}, \bibinfo {author} {\bibfnamefont {L.}~\bibnamefont {Wang}}, \bibinfo {author} {\bibfnamefont {K.}~\bibnamefont {Xu}}, \bibinfo {author} {\bibfnamefont {H.}~\bibnamefont {Li}}, \bibinfo {author} {\bibfnamefont {G.}~\bibnamefont
  {Wang}}, \bibinfo {author} {\bibfnamefont {K.}~\bibnamefont {Watanabe}}, \bibinfo {author} {\bibfnamefont {T.}~\bibnamefont {Taniguchi}}, \bibinfo {author} {\bibfnamefont {D.}~\bibnamefont {Qian}}, \bibinfo {author} {\bibfnamefont {J.}~\bibnamefont {Jia}}, \bibinfo {author} {\bibfnamefont {Q.}~\bibnamefont {Liang}}, \bibinfo {author} {\bibfnamefont {X.}~\bibnamefont {Wang}}, \bibinfo {author} {\bibfnamefont {W.}~\bibnamefont {Yang}}, \bibinfo {author} {\bibfnamefont {G.}~\bibnamefont {Zhang}}, \bibinfo {author} {\bibfnamefont {C.}~\bibnamefont {Jin}}, \bibinfo {author} {\bibfnamefont {W.}~\bibnamefont {Ouyang}},\ and\ \bibinfo {author} {\bibfnamefont {Z.}~\bibnamefont {Shi}},\ }\bibfield  {title} {\bibinfo {title} {{Homochiral carbon nanotube van der Waals crystals}},\ }\href {https://doi.org/10.1126/science.adu1756} {\bibfield  {journal} {\bibinfo  {journal} {Science}\ }\textbf {\bibinfo {volume} {387}},\ \bibinfo {pages} {1310} (\bibinfo {year} {2025})}\BibitemShut {NoStop}%
\bibitem [{\citenamefont {Le}\ \emph {et~al.}(2024{\natexlab{b}})\citenamefont {Le}, \citenamefont {Rodriguez-Lopez},\ and\ \citenamefont {Woods}}]{Le2024Phonon}%
  \BibitemOpen
  \bibfield  {author} {\bibinfo {author} {\bibfnamefont {D.-N.}\ \bibnamefont {Le}}, \bibinfo {author} {\bibfnamefont {P.}~\bibnamefont {Rodriguez-Lopez}},\ and\ \bibinfo {author} {\bibfnamefont {L.~M.}\ \bibnamefont {Woods}},\ }\bibfield  {title} {\bibinfo {title} {{Phonon-assisted Casimir interactions between piezoelectric materials}},\ }\href {https://doi.org/10.1038/s43246-024-00701-2} {\bibfield  {journal} {\bibinfo  {journal} {Commun. Mater.}\ }\textbf {\bibinfo {volume} {5}},\ \bibinfo {pages} {260} (\bibinfo {year} {2024}{\natexlab{b}})}\BibitemShut {NoStop}%
\bibitem [{\citenamefont {Ozaki}(2003)}]{Ozaki2003}%
  \BibitemOpen
  \bibfield  {author} {\bibinfo {author} {\bibfnamefont {T.}~\bibnamefont {Ozaki}},\ }\bibfield  {title} {\bibinfo {title} {{Variationally optimized atomic orbitals for large-scale electronic structures}},\ }\href {https://doi.org/10.1103/PhysRevB.67.155108} {\bibfield  {journal} {\bibinfo  {journal} {Phys. Rev. B}\ }\textbf {\bibinfo {volume} {67}},\ \bibinfo {pages} {155108} (\bibinfo {year} {2003})}\BibitemShut {NoStop}%
\bibitem [{\citenamefont {Bu{\v{c}}ko}\ \emph {et~al.}(2016)\citenamefont {Bu{\v{c}}ko}, \citenamefont {Leb{\`{e}}gue}, \citenamefont {Gould},\ and\ \citenamefont {{\'{A}}ngy{\'{a}}n}}]{Bucko2016}%
  \BibitemOpen
  \bibfield  {author} {\bibinfo {author} {\bibfnamefont {T.}~\bibnamefont {Bu{\v{c}}ko}}, \bibinfo {author} {\bibfnamefont {S.}~\bibnamefont {Leb{\`{e}}gue}}, \bibinfo {author} {\bibfnamefont {T.}~\bibnamefont {Gould}},\ and\ \bibinfo {author} {\bibfnamefont {J.~G.}\ \bibnamefont {{\'{A}}ngy{\'{a}}n}},\ }\bibfield  {title} {\bibinfo {title} {{Many-body dispersion corrections for periodic systems: an efficient reciprocal space implementation}},\ }\href {https://doi.org/10.1088/0953-8984/28/4/045201} {\bibfield  {journal} {\bibinfo  {journal} {J. Phys. Condens. Matter}\ }\textbf {\bibinfo {volume} {28}},\ \bibinfo {pages} {045201} (\bibinfo {year} {2016})}\BibitemShut {NoStop}%
\bibitem [{\citenamefont {Hermann}\ \emph {et~al.}(2023)\citenamefont {Hermann}, \citenamefont {St{\"{o}}hr}, \citenamefont {G{\'{o}}ger}, \citenamefont {Chaudhuri}, \citenamefont {Aradi}, \citenamefont {Maurer},\ and\ \citenamefont {Tkatchenko}}]{Hermann2023}%
  \BibitemOpen
  \bibfield  {author} {\bibinfo {author} {\bibfnamefont {J.}~\bibnamefont {Hermann}}, \bibinfo {author} {\bibfnamefont {M.}~\bibnamefont {St{\"{o}}hr}}, \bibinfo {author} {\bibfnamefont {S.}~\bibnamefont {G{\'{o}}ger}}, \bibinfo {author} {\bibfnamefont {S.}~\bibnamefont {Chaudhuri}}, \bibinfo {author} {\bibfnamefont {B.}~\bibnamefont {Aradi}}, \bibinfo {author} {\bibfnamefont {R.~J.}\ \bibnamefont {Maurer}},\ and\ \bibinfo {author} {\bibfnamefont {A.}~\bibnamefont {Tkatchenko}},\ }\bibfield  {title} {\bibinfo {title} {{libMBD: A general-purpose package for scalable quantum many-body dispersion calculations}},\ }\href {https://doi.org/10.1063/5.0170972} {\bibfield  {journal} {\bibinfo  {journal} {J. Chem. Phys.}\ }\textbf {\bibinfo {volume} {159}} (\bibinfo {year} {2023})}\BibitemShut {NoStop}%
\end{thebibliography}%

\end{document}